\documentclass[twocolumn]{aastex62}

\usepackage{graphicx}
\usepackage{float}
\usepackage{multirow}
\usepackage{color}
\usepackage[colorinlistoftodos]{todonotes}

\accepted{to ApJ, November 2019}
\begin{document}
	\title{Towards a more complex description of chemical profiles in exoplanet retrievals: A 2-layer parameterisation}
	
	\correspondingauthor{Q. Changeat}
	\email{quentin.changeat.18@ucl.ac.uk}
	\author[0000-0001-6516-4493]{Q. Changeat}
	\affil{Department of Physics and Astronomy \\
		University College London \\
		Gower Street,WC1E 6BT London, United Kingdom}
	\author[0000-0002-5494-3237]{B. Edwards}
	\affil{Department of Physics and Astronomy \\
		University College London \\
		Gower Street,WC1E 6BT London, United Kingdom}
	\author[0000-0002-4205-5267]{I.P. Waldmann}
	\affil{Department of Physics and Astronomy \\
		University College London \\
		Gower Street,WC1E 6BT London, United Kingdom}
	\author[0000-0001-6058-6654]{G. Tinetti}
	\affil{Department of Physics and Astronomy \\
		University College London \\
		Gower Street,WC1E 6BT London, United Kingdom}

\begin{abstract}

State of the art spectral retrieval models of exoplanet atmospheres assume constant chemical profiles with altitude. This assumption is justified by the information content of current datasets which do not allow, in most cases, for the molecular abundances as a function of pressure to be constrained.

In the context of the next generation of telescopes, a more accurate description of chemical profiles may become crucial to interpret observations and gain new insights into atmospheric physics. We explore here the possibility of retrieving pressure-dependent chemical profiles from transit spectra, without injecting any priors from theoretical chemical models in our retrievals. The ``2-layer'' parameterisation presented here allows for the independent extraction of molecular abundances above and below a certain atmospheric pressure.

By simulating various cases, we demonstrate that this evolution from constant chemical abundances is justified by the information content of spectra provided by future space instruments. Comparisons with traditional retrieval models show that assumptions made on chemical profiles may significantly impact retrieved parameters, such as the atmospheric temperature, and justify the attention we give here to this issue.

We find that the 2-layer retrieval accurately captures discontinuities in the vertical chemical profiles, which could be caused by disequilibrium processes -- such as photo-chemistry -- or the presence of clouds/hazes. The 2-layer retrieval could also help to constrain the composition of clouds and hazes by exploring the correlation between the chemical changes in the gaseous phase and the pressure at which the condensed phase occurs.

The 2-layer retrieval presented here therefore represents an important step forward in our ability to constrain theoretical chemical models and cloud/haze composition from the analysis of future observations.

\end{abstract}

\section{INTRODUCTION}
In the past years an increasing number of exoplanetary atmospheres have been characterised with space and ground-based observatories. Ultraviolet, optical and infrared spectra, recorded through transit, eclipse, high-dispersion and direct imaging, have offered a glimpse of the atmospheric structure and composition of exotic worlds orbiting other stars. 
In most cases the data available are sparse and therefore their interpretation is rarely unique. To explore the degeneracy,  reliability and correlations among the atmospheric parameters extracted from the data, the past decade has seen a surge in spectral retrieval models developed by many teams 
\citep[e.g.][]{Terrile,irwin2008,Madhu_retrieval_method,chimera,Waldmann_taurex1,cubillos2016,helios,ATMO,hydra}.

Most current spectral retrieval models assume constant or simplified atmospheric thermal profiles. Additionally,  chemical profiles which are constant with altitude are assumed  \citep[e.g.][]{MacDonald_HD209, Tsiaras_pop_study_trans,Pinhas_ten_HJ_clouds}. In these  models, the mixing ratio of each individual molecule is fully determined by a single free parameter. So far, this  approach has been successful due to the relatively poor quality of the input data from space and ground-based instruments. Given the low signal to noise, spectral resolution and the narrow wavelength coverage, current data cannot be used to constrain more complex models. However, the next generation of telescopes coming online in the next decade, will demand more complex retrievals to extract all the information content embedded in  the data. In the context of NASA-JWST \citep{Bean_JWST}, ESA-ARIEL \citep{Tinetti_ariel} and other facilities from ground and space (e.g. E-ELT \citep{brandl}, Twinkle \citep{Edwards_twinkle}), the higher resolution, signal-to-noise ratio (SNR) and  broader wavelength range will allow for less abundant trace gases and refined thermal profiles to be captured. For instance, \cite{Rocchetto_biais_JWST}
have demonstrated that the assumption of constant atmospheric thermal profiles will be inadequate to interpret correctly future better-quality transit spectra recorded from space. Additionally these new instruments may be sensitive enough to constrain non-constant chemical profiles.

The need for an increased chemical complexity is sometimes addressed in the literature through additional constraints in the retrievals from dynamical and chemical models  (``hybrid'' models). This method is already widely explored in retrievals aiming at constraining the thermal profiles, e.g. the Guillot model \citep{Guillot_TP_model} and other 2-stream approximations \citep{Heng_2stream,Malik_Helios}. This strategy allows for more complex thermal profiles to be considered, while limiting the number of free parameters. Similarly, equilibrium and disequilibrium chemical models may be used to constrain chemical profiles. \cite{Agundez_2dchemical_HD209_HD189} showed, with their 2D chemical model of HD\,209458\,b and HD\,189733\,b, that accurate parameterisations of exoplanetary atmospheres could be extremely complex.  Interesting alternatives combine both physical/chemical models and free parameters such as the model adopted by \cite{Madhu_retrieval_method}, where the chemical profiles are computed in equilibrium and multiplied by a factor to account for potential departures from the equilibrium.

The hybrid retrieval models  have, however, two major disadvantages. First, the forward model requires significant computing time to ensure convergence of the chemical/dynamical modules, which becomes even longer if used for retrievals. More fundamentally, they imply assumptions on the state of the planet and its physical/chemical behaviour. As the physics of such systems can be extremely complex and far from any environment we know in the Solar System, the selection of a particular model may lead to results biased by preconception. For instance, \cite{venot_chem_HJ} proposed a disequilibrium model adapted for hot-Jupiters and found significant differences when comparing with other models such as the equilibrium ones. This result highlights the issue of assuming a particular physics as a prior in inverse models, when our knowledge of exoplanetary atmospheres is still in an early phase. At least until our knowledge of these exotic worlds has progressed substantially, the results obtained by spectral retrievals should be kept independent from ab initio dynamical and chemical models, and used instead to constrain/validate some aspects of said models. 

The approach taken here is to increase the number of free variables for each molecular species considered. Applying this approach to currently available data is not justifiable as it would simply increase the degeneracy of the retrieved solutions. By contrast, attempts to use models of inadequate complexity to analyse spectra observed by next generation facilities are likely to provide incomplete pictures and misleading results. This paper explores the importance of moving towards a more complete description of chemical profiles through the analysis of  simulated transit data from JWST, ARIEL and other future telescopes. In that context, we consider the example of a 2-layer parametrisation with 3 degrees of freedom.

Section 2 presents the 2-layer approach and describes the methodology adopted. A validation of the method using simple cases is then reported in Section 3, followed by specific examples of exoplanetary atmospheres in section 4. Section 5 discusses the model's strengths and limitations. 

\section{METHODOLOGY}

\subsection{Overview and key assumptions}

This work focuses on retrievals of transit spectra, so that, for simplicity, the thermal profile can be assumed isothermal in some benchmark cases -- note that this assumption is abandoned in Section 4 to ensure our simulations are as realistic as possible. In eclipse spectroscopy, the thermal gradients and the chemical profiles are always entangled, making it a more complex case which will be considered in a separate paper.

The 2-layer parametrisation has been adopted for its simplicity and because it does not rely on external physical assumptions which, as previously described, could bias the results of the retrieval. While our model is clearly not representative of all real atmospheres, it allows us to consider a departure from the constant mixing ratios case.

Both the forward radiative transfer models and the inverse models (spectral retrievals) are based on the open-source TauREx from \cite{Waldmann_taurex2} and \cite{Waldmann_taurex1}, which has been modified for the purpose of this study. TauREx is a fully Bayesian radiative transfer and retrieval framework which encompasses molecular line-lists from the Exomol project \citep{Tennyson_exomol}, HITEMP \citep{rothman} and HITRAN \citep{gordon}. The complete list of opacity used in this paper can be found in Table \ref{fig_references}. The public version of TauREx\footnote{\url{https://github.com/ucl-exoplanets/TauREx_public}} is able to retrieve chemical composition of exoplanets by assuming constant abundances with altitude. It can also simulate atmospheres in equilibrium.

\begin{table}
\centering

$\begin{array}{|c|c|}
\mbox{Opacity} & \mbox{References}  \\
\hline
\mbox{H}_2\mbox{-H}_2 & \mbox{\cite{abel_h2-h2}, \cite{fletcher_h2-h2}} \\
\mbox{H}_2\mbox{-He} & \mbox{\cite{abel_h2-he}} \\
\mbox{H}_2\mbox{O} & \mbox{\cite{barton_h2o}, \cite{polyansky_h2o}} \\
\mbox{CH}_4 & \mbox{\cite{hill_xsec}, \cite{exomol_ch4}} \\
\mbox{CO} & \mbox{\cite{li_co_2015}} \\
\mbox{CO}_2 & \mbox{ \cite{rothman_hitremp_2010}} \\
\mbox{TiO} & \mbox{ \cite{schwenke_tio_1998}} \\

\end{array}$
\caption{List of opacities used in this work}
\label{fig_references}
\end{table}

\begin{center}
\begin{figure}[h]
\includegraphics[scale=0.6]{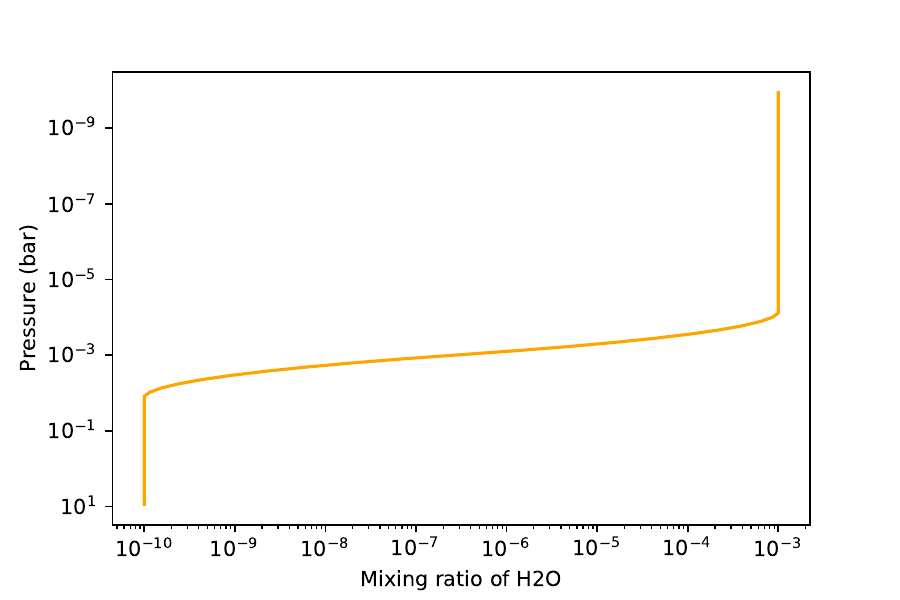}
\caption{
Example of a 2-layer chemical profile with H$_2$O. This profile can be used as input for forward simulations of exoplanet spectra, as well as for fitting data in retrievals. Here, the surface layer is depleted with a mixing ratio $X_S($H$_2$O$)$ of $10^{-10}$ and the top layer has a large quantity of H$_2$O, with $X_T($H$_2$O$) = 10^{-3}$. The separation pressure of the two layers is set to $P_I($H$_2$O$) = 10^{-3}$ bar and the transition is smoothed over 10\% of the atmosphere (10 layers).
}\label{fig:2layer_profile}
\end{figure}
\end{center}

Here we added the new 2-layer module to the code. The chemical parametrisation we used can be described by three variables: the surface/bottom abundance $X_S$, the top abundance $X_T$ and the pressure defining the separation of the two layers (Input Pressure Point $P_I$ for the forward model and Retrieved Pressure Point $P_R$). The chemical profile is linearly interpolated in log space -- smoothing over 10\% of the atmosphere -- to avoid a sharp transition in the profile. An example of a 2-layer chemical profile for water vapour is given in Figure \ref{fig:2layer_profile}. 
For all the tests reported in this paper, we follow the 3-step procedure detailed below.

\subsection{Step 1: generating high-resolution input spectra}

We start by using TauREx in forward mode and generate a high-resolution theoretical spectrum. In our models, we assumed a maximum pressure of $10$ bar, corresponding to the planet surface. The atmosphere is composed of the inactive gases H$_2$ and He, for which we set the ratio He/H$_2$ to 0.15 and add the considered active molecules (relative abundance defined by their mixing ratio). We consider collision induced absorption of the H$_2$-H$_2$ and H$_2$-He pairs and opacities induced by Rayleigh scattering \citep{cox_allen_rayleigh}. Throughout the paper, the molecular mixing ratios and profiles in the forward model are varied to create high resolution spectra for a wide range of compositions and cases. The planetary parameters have been set to the well known exoplanet HD\,209458\,b in Section 3 and  are listed in the Appendix. The adopted approach is compatible with any other set of planetary and atmospheric parameters and is applied in section 4 to two other simulated planets inspired by WASP-33\,b and GJ\,1214\,b.


\subsection{Step 2: convolution of the input spectra with the instrument response function}

\begin{center}
\begin{figure}[h]
\includegraphics[scale=0.4]{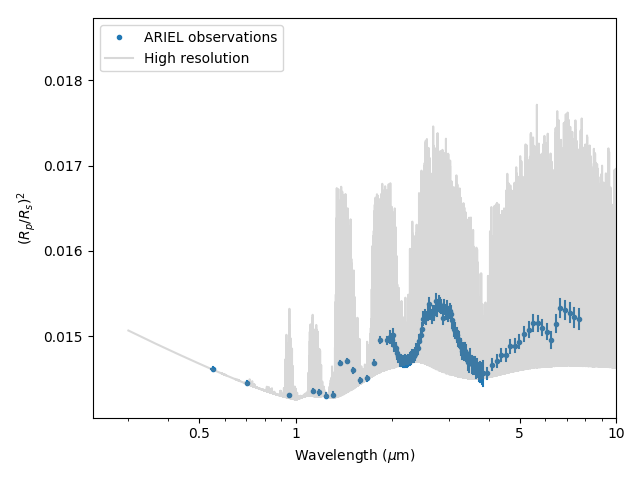}
\caption{
Example of forward models assuming a 2-layer chemical profile of H$_2$O as shown in Figure \ref{fig:2layer_profile}. Grey curve:  high-resolution theoretical spectrum obtained with TauREx at step 1 (sec. 2.2). Blue curve:  simulated ARIEL observations after having processed the theoretical spectrum in step 2 (sec. 2.3).
}\label{fig:Ariel_bin}
\end{figure}
\end{center}

High-resolution theoretical spectra obtained in sec. 2.2  are convolved with the instrument response function to simulate realistic observations. We use ArielRad \citep{mugnai_Arielrad} to provide realistic noise models for spectra obtained by ARIEL and chose planets that are in the current target list \cite{edwards_ariel}. In the case of JWST, we used the noise estimates for HD\,209458\,b presented in \cite{Rocchetto_biais_JWST}. 
An example of this process is shown in Figure \ref{fig:Ariel_bin} where both the high-resolution theoretical spectrum and the ARIEL-simulated case are presented.

\subsection{Step 3: Retrievals}

 We run TauREx in retrieval mode and use the spectra obtained in step 2 as input to the retrieval. We used the nested sampling algorithm Multinest from \cite{Feroz_multinest} with 1500 live points and a log likelihood tolerance of 0.5. The retrieved parameters include our chemical setup (3 variables per chemical species), the isothermal temperature value and the planet radius. We therefore have a minimum of 5 free parameters that we attempt to retrieve. In the case where the mixing ratios were assumed constant with altitude (1-layer forward model), the Retrieved Pressure Point has been fixed, so that we have only 2 free variables per chemical species or a minimum of 4 free parameters. In our retrieval scheme, we use uniform priors for all the free parameters. In all our retrievals, chemical abundances are allowed to explore the bounds $10^{-12}$ to $10^{-1}$. For the Retrieved Pressure Point, we allow the bounds from $10^{-1}$ to $10^{-4}$ bar for the Section 3. In Section 4, since we investigate less ad-hoc situations, we allow the pressure to explore $10^{-1}$ bar to $10^{-7}$ bar. For the isothermal temperature retrievals, the priors span $\pm$ 30 percent of the ground truth value. In section 4, since we investigate more realistic examples, we retrieve a 3-point temperature profile (\cite{Waldmann_taurex2}). The atmospheric parameters used to generate the theoretical spectrum in step 1 are the ground truth. By comparing the posteriors obtained by the retrieval to the ground truth, we can test the reliability and accuracy of the retrieval process. Furthermore, by using the predicted performances of  JWST and ARIEL, we can quantify the expected information content of future data, with a view to assessing the ability to probe the chemical complexity of exoplanet atmospheres. For each retrieval, we state the Nested Sampling Log-Evidence. Bayes factor B \citep{jeffreys1998_bayesfactor}, which is the ratio of the evidences of two competing models ($E_1$ and $E_2$), allows to compare models against each other. In practice, the table in \cite{Kass1995bayes} gives an interpretation of $log(B) = \Delta log(E) = log(E_2) - log(E_1)$: 
\begin{table}
\centering

$\begin{array}{|c|c|}
\mbox{$log(B)$} & Interpretation  \\
\hline
\mbox{0 to 0.5} & No\ Evidence \\
\mbox{0.5 to 1} & Some\ Evidence \\
\mbox{1 to 2} & Strong\ Evidence  \\
\mbox{$>$ 2} & Decisive  \end{array}$
\caption{Interpretation of the Bayes ratio \citep{Kass1995bayes}}
\end{table}

By applying the 3-step methodology, a number of cases are simulated. Firstly, we verify that the 2-layer retrieval  is able to recover the more basic 1-layer input  (i.e. a constant chemical profile). We then investigate the ``retrievability'' of the 2-layer input spectrum by a 2-layer retrieval in the case of  JWST and ARIEL observations. Finally, we explore the advantage of using a 2-layer approach by comparing how a 2-layer input spectrum is recovered by both 1- and 2-layer  retrievals. 

\subsection{Testing the 2-layer approach: Retrieval of a 1-layer input spectrum using the 2-layer parametrisation}
As a sanity check, we test that the more complex 2-layer model can indeed recognise the simple case of constant chemistry. A 1-layer simulated spectrum  is generated and we attempt to recover the solution using the 2-layer model. Here, the retrieval of the pressure point ($P_R$) is disabled as this parameter introduces intrinsic degeneracy in the specific case of constant chemistry. Here, the goal being to illustrate that our 2-layer model returns the expected solution when tested on the 1-layer forward model, the behaviour of the retrieval when the Retrieved Pressure Point is activated is discussed in section 4. As any value for this point would work,  we arbitrarily choose to set it at $P_I = 10^{-1.3}$ bar.

\subsection{Retrieval of a 2-layer input spectrum as observed by  JWST and ARIEL}
We   study an exoplanet exhibiting noticeable chemical modulations with altitude. This case can be simulated by using a 2-layer profile as input. We present the particular case of an input H$_2$O profile with 2 layers separating at $P_I($H$_2$O$) = 10^{-2}$ bar. The input H$_2$O surface layer is set with a mixing ratio of $X_S($H$_2$O$) = 10^{-3}$ and the top layer contains $X_T($H$_2$O$) = 10^{-5}$.
The input spectrum is simulated at high-resolution and observations are reproduced by convolving the theoretical spectrum to the instrument response function  of  JWST and ARIEL.

\subsection{Comparison of the 1-layer and  2-layer retrievals.}
By comparing the results obtained with  the 1-layer and 2-layer retrievals, we aim to illustrate issues that may occur when performing a retrieval with a model of inappropriate complexity. Therefore, we simulate planetary atmospheres with  2-layer chemical profiles and analyse the results if the retrieval is performed with a 1-layer chemical approach. For this test, we use ARIEL simulations  to illustrate our results. 
In particular,
two main issues could occur and need to be tested: 
\begin{enumerate}
\item The observed spectrum cannot be explained using the 1-layer
retrieval, as the best solution retrieved does not fit the data. 
\item The 1-layer retrieval manages to achieve a ``good'' fit but the retrieved parameters are wrong compared to the ground-truth. This issue is more subtle as there is little evidence and no direct way to spot the error.  
\end{enumerate}
These two points can be tested by considering the following examples. For the former, we assume an atmosphere with a single CH$_4$ profile with a surface layer of $X_S($CH$_4) = 10^{-5}$ up to $P_I($CH$_4) = 10^{-2}$ bar and  $X_T($CH$_4) = 10^{-10}$ above that pressure, corresponding to a depleted layer. For the latter,  we simulate a single H$_2$O  profile where the planet contains $X_S($H$_2$O$) = 10^{-10}$ up to $10^{-2}$ bar and  the mixing ratio is $X_T($H$_2$O$)= 10^{-3}$ for lower pressures. 

\section{RESULTS}
\subsection{Testing the 2-layer approach: Retrieval of a 1-layer input spectrum using the 2-layer parametrisation}
\label{const with 2 layer}
\begin{figure}[H]
    \centering
    \includegraphics[width=0.45\textwidth]{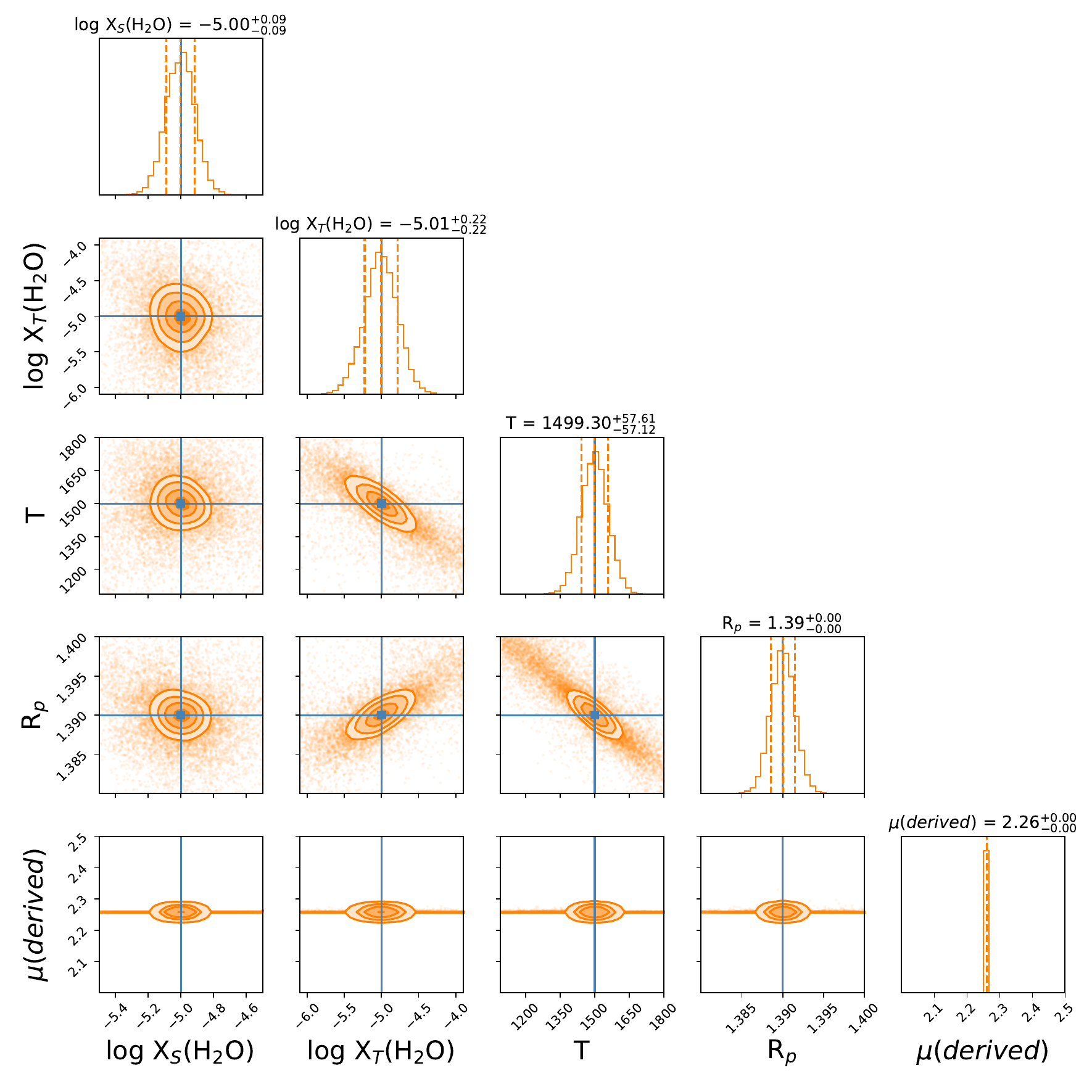}
    \caption{Posterior distributions of a 1-layer input atmosphere  retrieved using the 2-layer model. The input spectrum was generated by assuming a constant profile for H$_2$O with a mixing ratio of $10^{-5}$. In this example, the Retrieved Pressure Point is disabled and arbitrarily set at $P_I = 10^{-1.3}$ bar. For each free parameter, we report the mean and 1-sigma iso-likelihood levels with the dashed lines. The retrieved values match the input values within the retrieved uncertainties.} \label{fig:cst_2layer_posteriors}
\end{figure}

 The retrieved posterior distributions for an input spectrum generated with  1-layer parametrisation  with a single species, H$_2$O, is presented in Figure \ref{fig:cst_2layer_posteriors}. In orange, we show the retrieved posterior distribution of the parameters, while the true value (when available) is marked in blue.
 The mixing ratio of H$_2$O used for this example was $10^{-5}$. The 2-layer model successfully retrieved the same abundance for both layers. This result matches the single input parameter and confirms that the 2-layer parametrisation can recover the 1-layer input. This example showcases a situation where the complexity of the retrieval model is higher than the input.

\subsection{Retrieval of a 2-layer input spectrum as observed by  JWST and ARIEL}
\begin{center}
\begin{figure}[h]
\centering

    \includegraphics[width=0.45\textwidth]{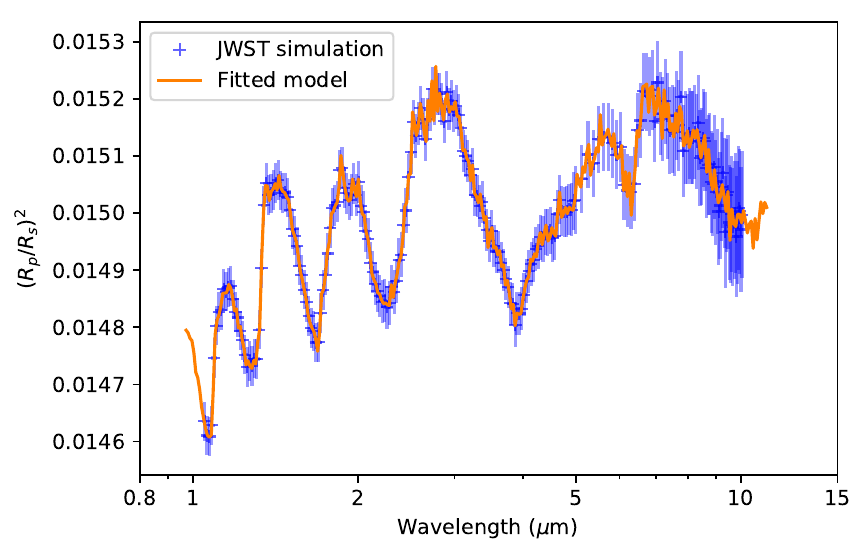}

    \includegraphics[width=0.45\textwidth]{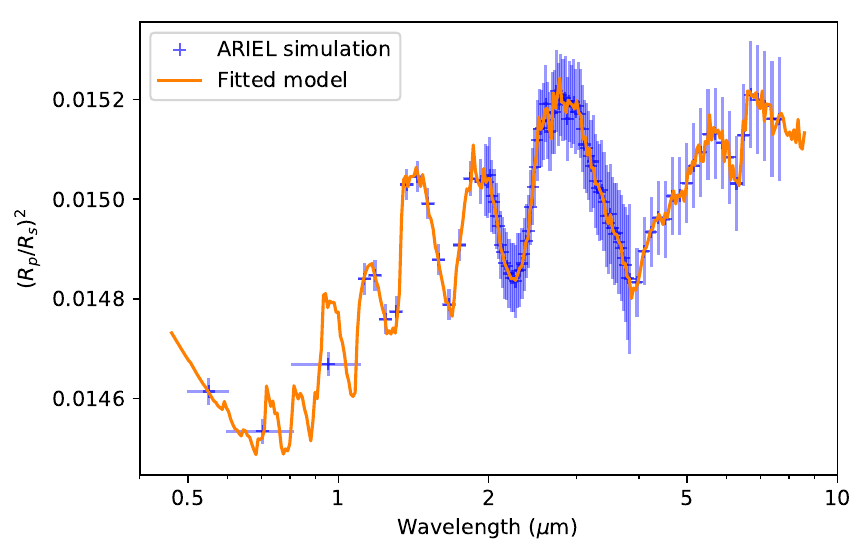}
\caption{
Retrieved spectra for a Hot-Jupiter similar to HD\,209458\,b which presents a 2-layer H$_2$O chemical profile. Case without Gaussian scatter added to the data points. Top: JWST simulated performance ($log(E) = 2076$); Bottom: ARIEL simulated performance ($log(E) = 883$).
}\label{fig:H2On_retrievals}
\end{figure}
\end{center}

\begin{center}
\begin{figure}[h]
\centering

    \includegraphics[width=0.45\textwidth]{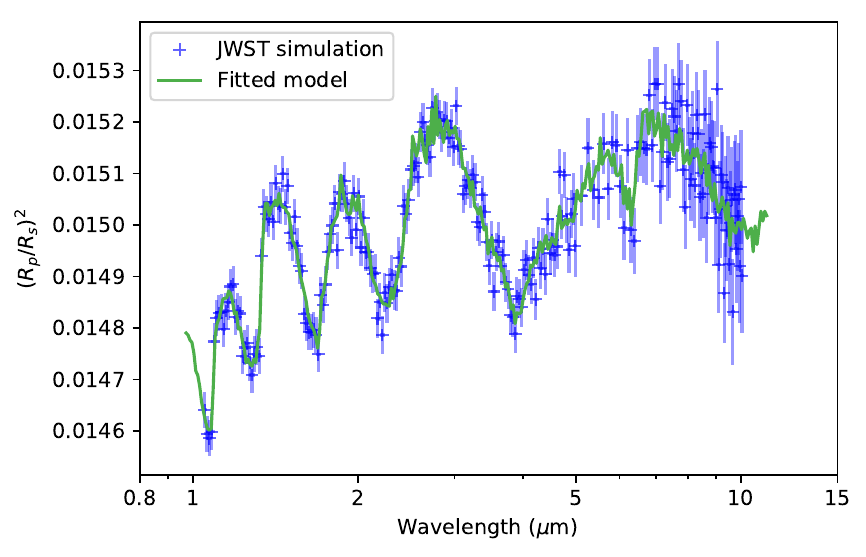}

    \includegraphics[width=0.45\textwidth]{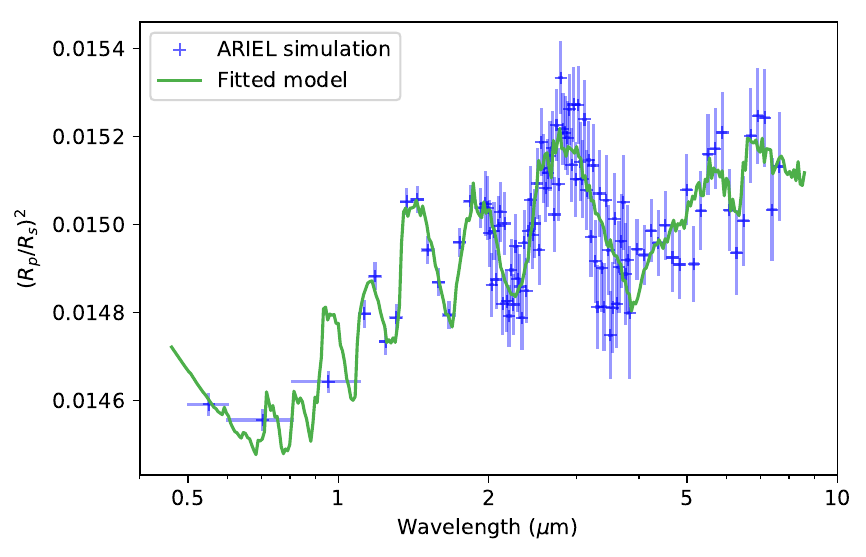}
\caption{
Retrieved spectra for a Hot-Jupiter similar to HD\,209458\,b with a 2-layer H$_2$O chemical profile. Gaussian scatter was added to the mean data points. Top: JWST simulated performance ($log(E) = 1985$); Bottom: ARIEL simulated performance ($log(E) = 828$).
}\label{fig:H2On_retrievals_noise}
\end{figure}
\end{center}

\begin{center}
\begin{figure}[h]
\centering
    \includegraphics[width=0.48\textwidth]{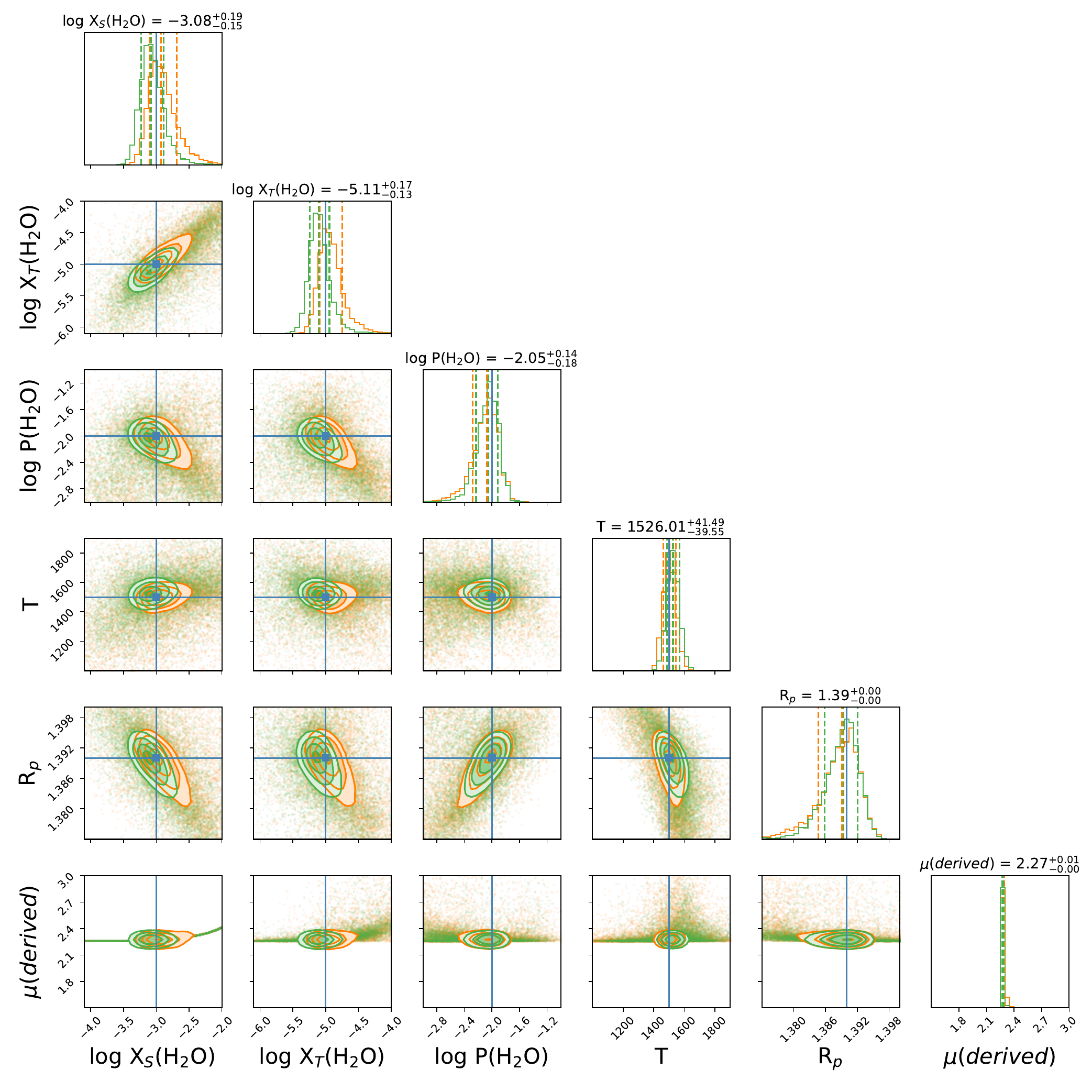}
\centering
    \includegraphics[width=0.48\textwidth]{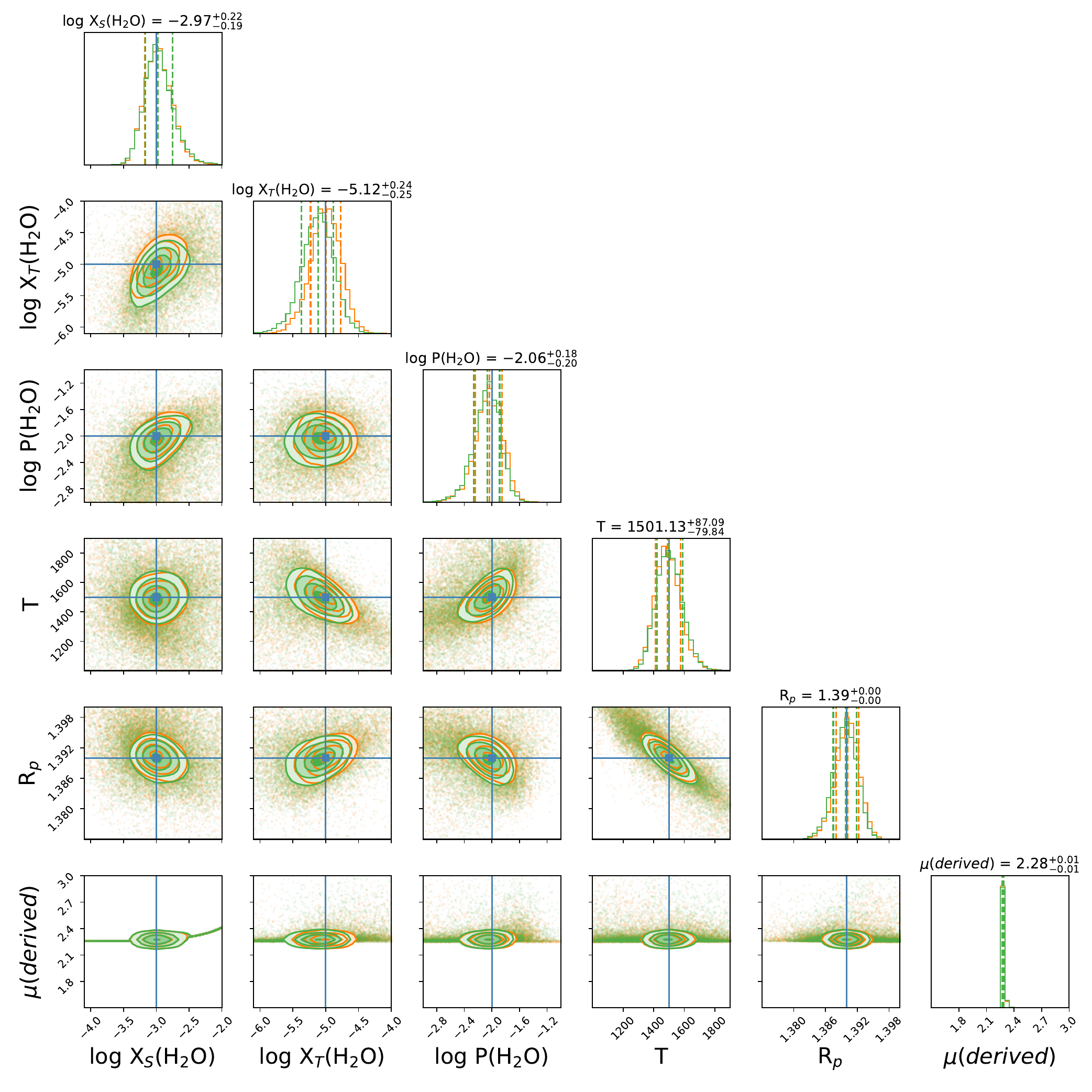}
\caption{
Posterior distributions of the  JWST and ARIEL simulations (Fig. \ref{fig:H2On_retrievals} and Fig. \ref{fig:H2On_retrievals_noise}) for a Hot-Jupiter similar to HD\,209458\,b, which presents a 2-layer H$_2$O chemical profile. The input atmosphere contains  $X_S($H$_2$O$) = 10^{-3}$ for the bottom layer and $X_T($H$_2$O$) = 10^{-5}$ for the top layer of the atmosphere. The layers separation is set at $P_I($H$_2$O$) = 10^{-2}$ bar and the retrieval is performed with the 2-layer model. In both the  JWST and ARIEL retrievals, all chemical variables are well retrieved. Top: JWST; Bottom: ARIEL. The orange posteriors corresponds to the case without noise, while the green ones are for the noisy cases.
}\label{fig:2layer_2layer_posteriors}
\end{figure}
\end{center}

The input H$_2$O surface layer was set as $X_S($H$_2$O$)= 10^{-3}$ and the top layer contained $X_T($H$_2$O$) = 10^{-5}$. 
These abundances result in strong features in the spectrum but  additional retrievals  show that similar conclusions can be obtained for mixing ratios down to $10^{-6}$ and for other molecules. The limits of the model are discussed in  section 5.  The planet parameters are retrieved for JWST and ARIEL using the 2-layer model. The best fitted spectra are presented in Figure \ref{fig:H2On_retrievals} for the two telescopes. We also run the same cases with the noised up spectra, where we applied a dispersion corresponding to the uncertainties (see Figure \ref{fig:H2On_retrievals_noise}). We show the posterior distributions of both cases for the JWST and ARIEL simulations in Figure \ref{fig:2layer_2layer_posteriors}.

The model is able to recognise the two layers in ARIEL and JWST simulations for the scattered and non-scattered cases. 


Typically, to simulate an observation instance, the flux per wavelength observed, $F_\lambda$, is drawn from a Gaussian distribution (in the limit of $N_{photons}$ being large) defined by its 1\,$\sigma$ error-bar and the ``noise-free" mean flux of the forward model, $\bar{F_\lambda}$. Here, we decided not to sample from this distribution and to adopt the "noise-free" mean for the following reason. In this publication we are interested in the intrinsic biases resulting from using over-simplified to more complex (1-layer vs 2-layer) chemical profiles. To study these biases, one can either generate thousands of noise-instances, $N_{instance}$ and average their retrieval results to obtain the underlying mean of the distribution, or avoid adding noise to $\bar{F_\lambda}$ in the first place. Given that in this case, all noise is Normally distributed and following the central-limit theorem (in the limit of large $N_{instance}$) both approaches are equivalent. \cite{Feng_retrieval_earthanalog} showed this to be the case and adopted the same rationale in their study. For the rest of the paper we therefore do not scatter our simulated spectra.

In the ARIEL case, we accurately retrieved both the surface layer ($X_S($H$_2$O)$ = 10^{-2.97}$) and the top layer ($X_T($H$_2$O$) = 10^{-5.12}$). The Retrieved Pressure Point also matched the input parameters ($P_R($H$_2$O$) = 10^{-2.06}$ bar). The same conclusions are reached for JWST. Additionally, in both simulations, the 2-layer retrievals  recovered the correct temperature of 1500K and radius (1.39 $R_J$) corresponding to the input. 

\subsection{Comparison between the 1-layer and  2-layer retrievals.}
\begin{center}
\begin{figure}[h]
\centering
    \includegraphics[width=0.49\textwidth]{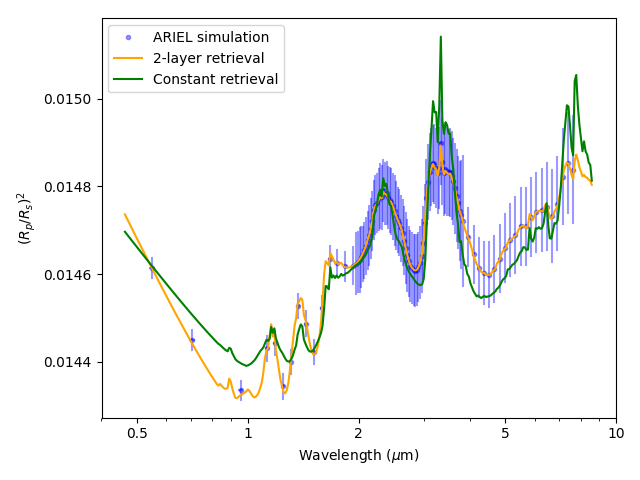}
\caption{
Observed input spectrum obtained with a 2-layer CH$_4$ profile  and retrieved spectrum obtained with a 1-layer  retrieval. This example showcases that the 1-layer retrieval is inadequate to interpret the data. The correct 2-layer retrieval is also shown. The Nested Sampling Global Evidence is $log(E) = 737$ for the 1-layer and $log(E) = 885$ for the 2-layer retrieval. This implies $log(B) = 148$, which is decisively in favour of the 2-layer scenario.
}\label{fig:CH4_2layer_cst_spectrum}
\end{figure}
\end{center}

We show here the results of the test where the input spectrum was generated assuming  CH$_4$ only with
$X_S($CH$_4) = 10^{-5}$ up to $P_I($CH$_4) = 10^{-2}$ bar and $X_T($CH$_4) = 10^{-10}$.
$X_T($CH$_4) = 10^{-10}$ does not produce any observable feature, so for this layer we expect to retrieve only an upper limit in the posteriors. In this example, the 1-layer retrieval has difficulties in fitting the observed spectrum, as shown in Figure \ref{fig:CH4_2layer_cst_spectrum}.
In this case, the 1-layer retrieval lacks flexibility, which leads to a poor fit of the spectrum. This is also backed up by the lower Nested Sampling Global Evidence for the 1-layer scenario: 737 for the 1-layer and 885 for the 2-layer ($\Delta log(E) = 148$). This example illustrates the need for a 2-layer retrieval.  

\begin{center}
\begin{figure}[h]
\centering
    \includegraphics[width=0.49\textwidth]{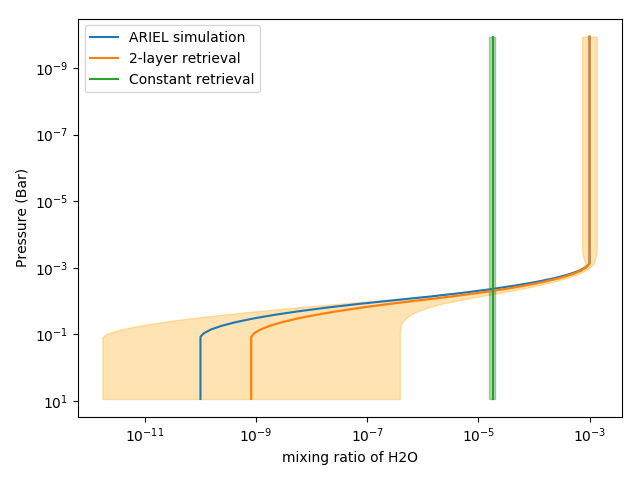}
    \includegraphics[width=0.49\textwidth]{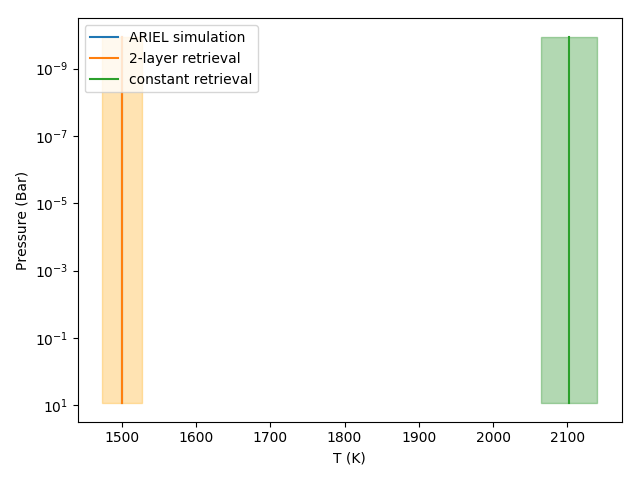}
\caption{
Chemical (top) and temperature (bottom) profiles for the input atmospheric model, the 2-layer and 1-layer retrievals. For the temperature, the 1-layer model is strongly biased. The input model temperature is not clearly visible as it overlaps with  the retrieved value of the 2-layer retrieval at 1500K.    
}\label{fig:H2O_2layer_vs_profiles}
\end{figure}
\end{center}

Concerning the test where the input spectrum was generated with H$_2$O only and assuming $X_S($H$_2$O$) = 10^{-10}$ and  $X_T($H$_2$O$) = 10^{-3}$ above $10^{-2}$ bar, both the 1-layer and 2-layer retrievals converged  to a solution {and gave a satisfactory fits of the input spectrum (See Figure \ref{fig:H2O_2layer_cst_spectrum}).} The posterior distributions are presented in Appendix, Figure \ref{fig:appendix_H2O}. Unsurprisingly, the 2-layer retrieval managed to recover the correct input parameters. However, while fitting the spectrum, significant differences appear for the 1-layer model in the retrieved parameters. The 1-layer retrieval tries to compensate the lack of flexibility in the chemical profile by increasing the temperature to 2100K {\
instead of the 1500K ground truth temperature.} The input chemical and thermal profiles for both retrievals are shown in Figure \ref{fig:H2O_2layer_vs_profiles}.

The retrieved temperature by the 1-layer retrieval is significantly off compared to the input, while the retrieved H$_2$O mixing ratio approximates the atmospheric average. 
This example illustrates well  the importance of exploring and understanding more complex chemical models in retrievals. Here the retrieved spectrum using the 1-layer approximation (Figure \ref{fig:H2O_2layer_cst_spectrum}) gives an acceptable fit while leading to a wrong solution, which is a serious issue. Small differences compared to the  observations are noticeable which, in this case, would still permit the selection of the 2-layer solution, provided that both retrievals are performed. More importantly, the correct solution can  be determined by comparing the Nested Sampling Global Log-Evidence of the retrieval. The 2-layer retrieval obtained a value of $log(E)=883$ while the 1-layer only had $log(E)=733$, indicating a clear preference for the 2-layer scenario (difference of $\Delta log(E) = 150$).

\begin{center}
\begin{figure}[h]
\centering
    \includegraphics[width=0.49\textwidth]{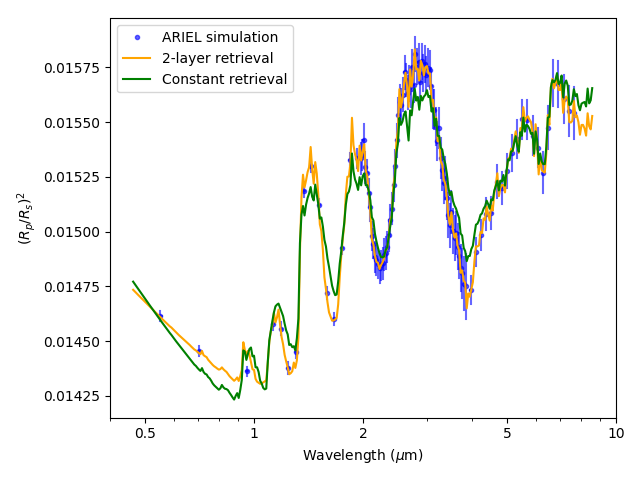}
\caption{
Observed spectrum generated with a 2-layer H$_2$O profile as input and best retrieved solutions obtained with a 1-layer and 2-layer retrievals. While the 2-layer retrieval captures better the observations, the differences with the 1-layer fit are relatively small. The Nested Sampling Global Evidence is $log(E) = 733$ for the 1-layer and $log(E) = 883$ for the 2-layer retrieval.
}\label{fig:H2O_2layer_cst_spectrum}
\end{figure}
\end{center}


\section{More realistic examples}

The previous sections demonstrated the theoretical possibility and, in some cases, the necessity of retrieving 2-layer chemical profiles in a number of select, simplified examples. Here we test the 2-layer approach by applying it to two cases inspired by GJ\,1214\,b and WASP-33\,b. Spectra and parameters used in this section should not be considered as the ``true'' values of these real planets: they are realistic scenarios inspired from examples of the literature, which are here used to explore advantages and limitations of the 2-layer approach. 

For instance, recent observations of the sub-Neptune GJ\,1214\,b in the near-IR unveiled a featureless spectrum, which could be caused by high-altitude hazes. Also, recent observations of the ultra hot-Jupiter WASP-33\,b point towards a thermal inversion and a significant amount of TiO, which could cause this inversion by acting as a strong absorber in the visible. We simulate the WASP-33 b case with a 2-layer input model, while in the GJ\,1214 b case we highlight the pertinence of our model by using the profiles from a disequilibrium chemistry model as input. We detail below all the assumptions we considered for our tests.

\subsection{An ultra Hot-Jupiter inspired by WASP-33\,b}

Current analyses of ground and space-based observations of WASP-33\,b suggest extreme temperatures reaching 3800\,K and a possible thermal inversion in the atmosphere \citep{Haynes_Wasp33b_spectrum_em, Nugroho_W33b}. TiO or VO, which are strong absorbers at short wavelengths, could very efficiently capture high-energy stellar photons  at the top of the atmosphere and cause the inversion  \citep{fortney,Spiegel_TiO_invertion}. In parallel, other observations have suggested the presence of TiO in Wasp-121\,b \citep{Evans_Wasp121b_spectrum_em} and Wasp-76\,b \citep{Tsiaras_pop_study_trans}. 

Here we investigate this process by attempting to detect a TiO layer in the upper atmosphere of a simulated planet resembling WASP-33\,b. Our input model includes only two molecules: H$_2$O and TiO. The simulation consists of a constant mixing ratio of $10^{-4}$ for H$_2$O and an inverted temperature-pressure (TP) profile from 2800K to 3700K, which is inspired by \cite{Haynes_Wasp33b_spectrum_em}. For the temperature-pressure profile we used a 3-point model \citep{Waldmann_taurex2}. This model interpolates a smooth TP profile using 5 free parameters, i.e. surface temperature and two temperature-pressure points. The temperature variations allow us to explore the possibility of retrieving both thermal and chemical parametric profiles at the same time. \cite{Rocchetto_biais_JWST} have shown that non-isothermal profiles could introduce a bias in JWST observations. This is also expected to be true for ARIEL and we can investigate it as a side result of this work. For the retrieval, we explore uniform priors on the pressure bounds $10^{-2}$ - $10^{1}$ bar for the point 1 and $10^{-5}$ - $10^{-1}$ bar for the point 2. The temperature bounds are the same for all 3 retrieved points and cover the a range 30 percent lower/higher than the input min/max temperatures (1960K - 4690K). For a real observation, these priors could be informed by the knowledge of the equilibrium temperature and the physics of the atmosphere. To simulate a stratospheric TiO layer, we assumed abundances of $X_T($TiO$) = 10^{-4}$ for the top layer (down to $P_I = 10^{-4}$ bar) and $X_S($TiO$) = 10^{-7}$ at the surface. The spectrum, as well as the temperature and chemical profiles, are presented in Figure\,\ref{fig:W33b_retrieval} while the full posterior distribution is available in the Appendix, Figure \ref{fig:appendinx_Wasp33b}.

\begin{center}
\begin{figure}[h]
\centering
    \includegraphics[width=0.45\textwidth]{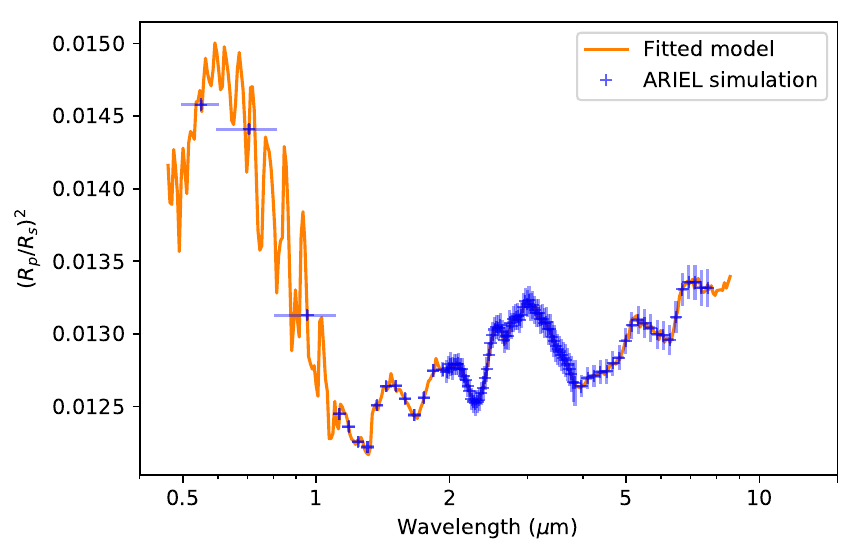}
    \includegraphics[width=0.45\textwidth]{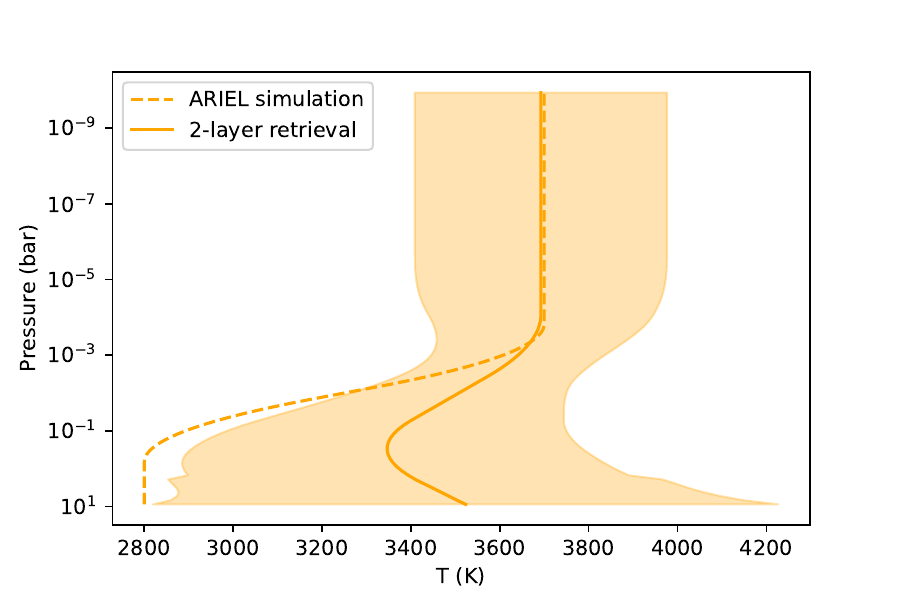}
    \includegraphics[width=0.45\textwidth]{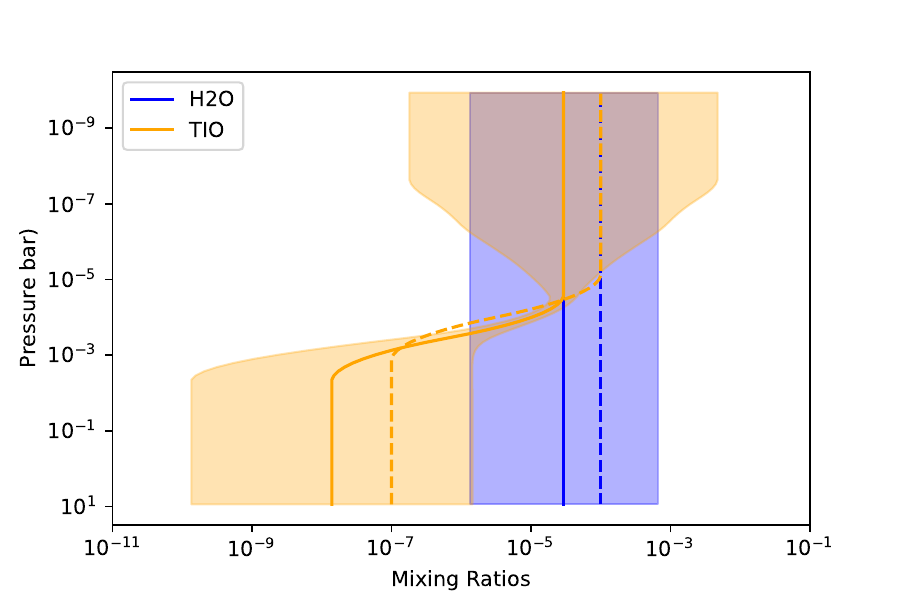}
\caption{
Outcome of the WASP-33\,b retrieval simulations. Top:  fitted spectrum; Middle: retrieved temperature profile; Bottom: retrieved chemical profiles of H$_2$O and TiO. The dashed lines correspond to the input values assumed in the forward model while the solid lines indicate the retrieved profiles. For this run log(E) = 880.
}\label{fig:W33b_retrieval}
\end{figure}
\end{center}

These results demonstrate the possibility of accurately retrieving the vertical distribution of the TiO layer. In particular, the TiO profile is well constrained between $10^{-6}$ bar and $10^{-3}$ bar as a result of the strong features between 0.4 $\mu m$ and 1 $\mu m$. The shape of the thermal profile is also correctly retrieved, although with a larger uncertainty at higher pressures as shown by the posterior distribution. The retrievability of the thermal and chemical profiles at the same time indicates that retrievals of future transit spectra should take these two effects into account. The flexibility of the 2-layer approach allows for the confirmation (or rejection) of potential correlations between molecules/condensates and thermal inversions. This is an important application of the 2-layer approach.

\subsection{A warm sub-Neptune inspired by GJ\,1214\,b}

As previously mentioned, GJ\,1214\,b is a sub-Neptune with a relatively flat spectrum in the visible and near-infrared. Multiple explanations for the lack of features have been proposed \citep{Miller_GJ1214_clouds, Morley_GJ1214_clouds, Kreidberg_GJ1214b_clouds}:
\begin{itemize}
\item The planet could have an atmosphere heavier than hydrogen, such as a water dominated atmosphere. \
\item The atmosphere could be hydrogen dominated with opaque, high altitude clouds (e.g. KCl or ZnS). 
\item The planet could have hydrocarbon hazes in the upper atmosphere.  
\end{itemize}

Here we investigate the retrievability of the third scenario as an example of 2-layer chemistry. The photo-chemical hazes could be similar to those found in the atmosphere of Saturn's moon Titan. CH$_4$ in the upper atmosphere is photolysed by radiation, creating hydrocarbon hazes that are opaque in the near infrared. This scenario implies a significant CH$_4$ abundance at the surface of the atmosphere and a sharp decline in the upper atmosphere. For our input forward model, we used the chemical profiles of the four molecules H$_2$O, CH$_4$, CO$_2$ and CO published by \cite{Miller_GJ1214_clouds} in the case of solar abundance and $Kzz = 10^6 $ $cm^2s^{-1}$. The profiles are retrieved using two scenarios: A hybrid one with 2-layer chemistry for H$_2$O and CH$_4$ and the 1-layer for CO$_2$ and CO; and a fully 1-layer chemistry with all 4 molecules retrieved using constant profiles. For the hybrid case, the abundances of CO$_2$ and CO were indeed too low to retrieve the 2-layer profile. The other input parameters for the planet are described in the Appendix. 
We add hydrocarbon hazes layer adopting the model described in \cite{Lee_haze_model}. The clouds are treated as additional opacity $\sigma_c$ for each layer of size $\Delta z$:
\begin{equation}
    \sigma_c = Q_{ext} \pi R_{clouds}^2 \chi_{clouds} \Delta z.
\end{equation}
Where $R_{clouds}$ is the size of the cloud particles and $\chi_{clouds}$ is the cloud number density. Here the extinction efficiency $Q_{ext}$ has been defined as:
\begin{equation}
    Q_{ext} = \frac{5}{Q_0 x^{-4}+x^{0.2}},
\end{equation}
and the cloud size parameter $x$ is:
\begin{equation}
    x = \frac{2 \pi R_{clouds}}{\lambda}.
\end{equation}

\begin{figure*}
\centering
    \includegraphics[width=0.45\textwidth]{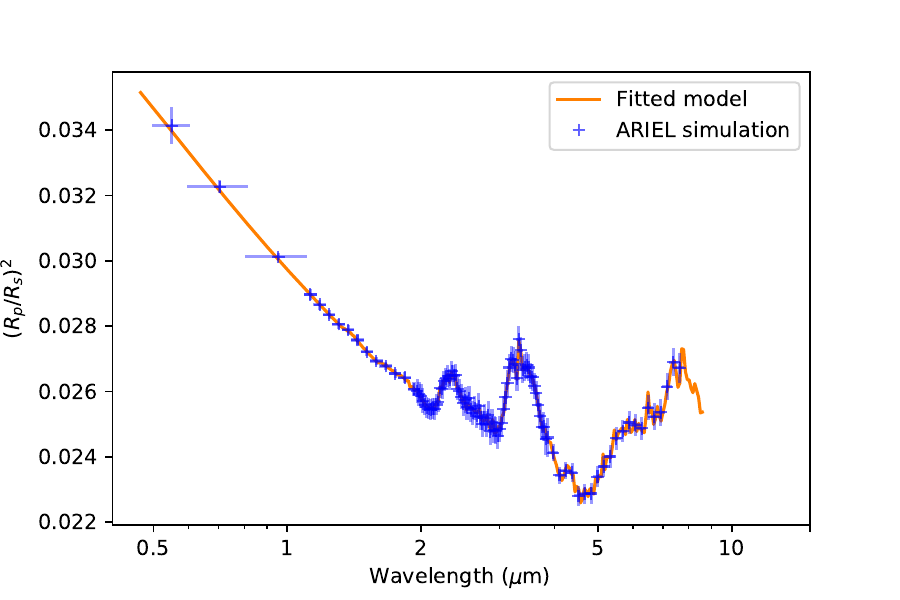}
    \includegraphics[width=0.45\textwidth]{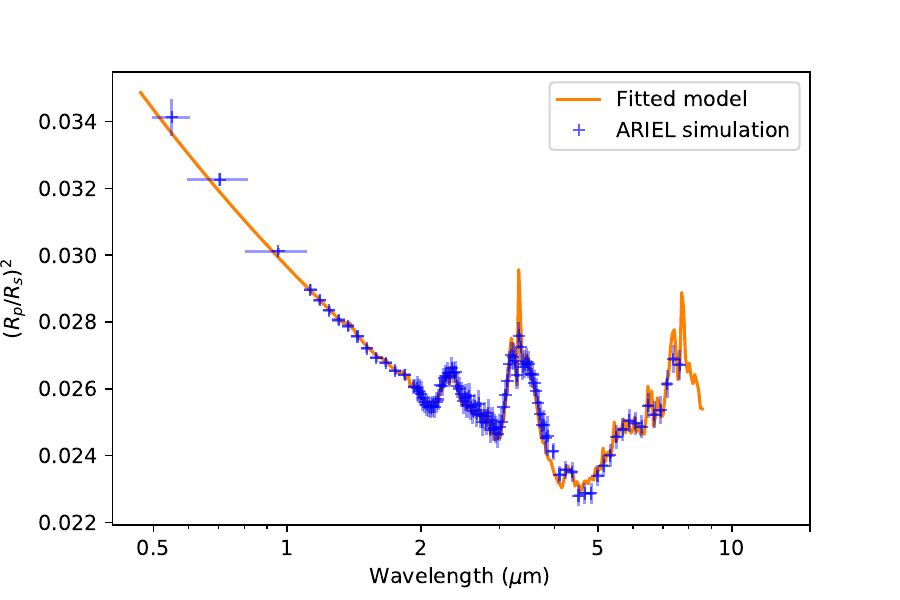}
    \includegraphics[width=0.45\textwidth]{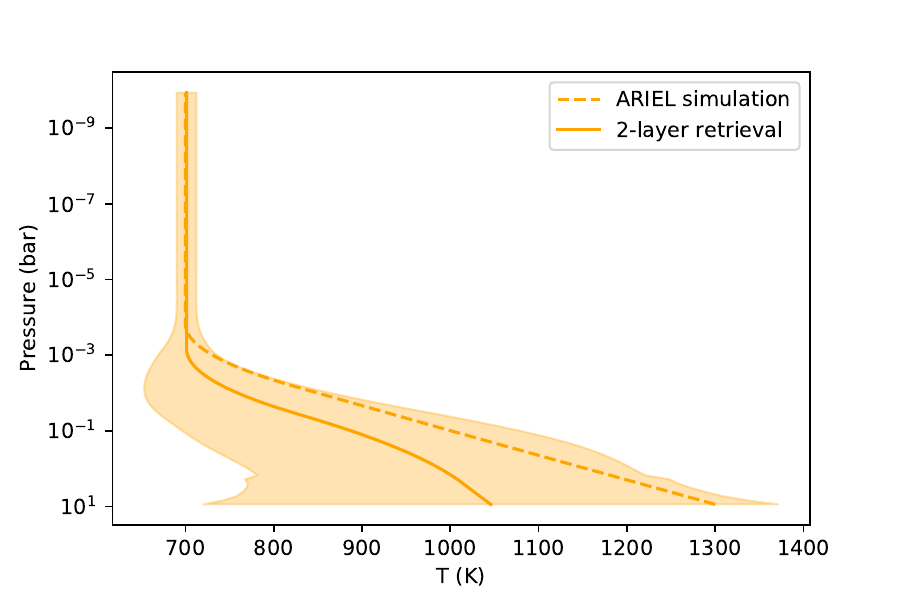}
    \includegraphics[width=0.45\textwidth]{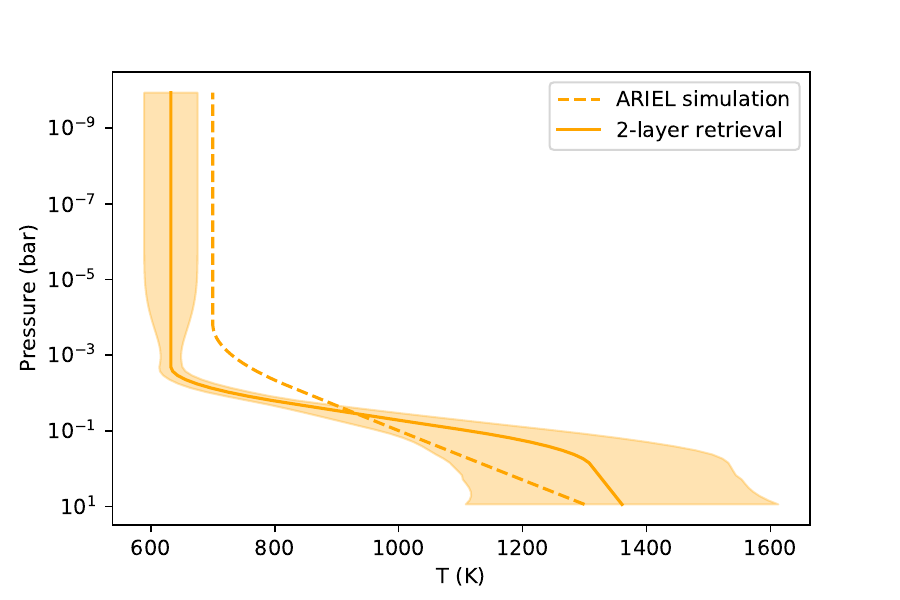}
    \includegraphics[width=0.45\textwidth]{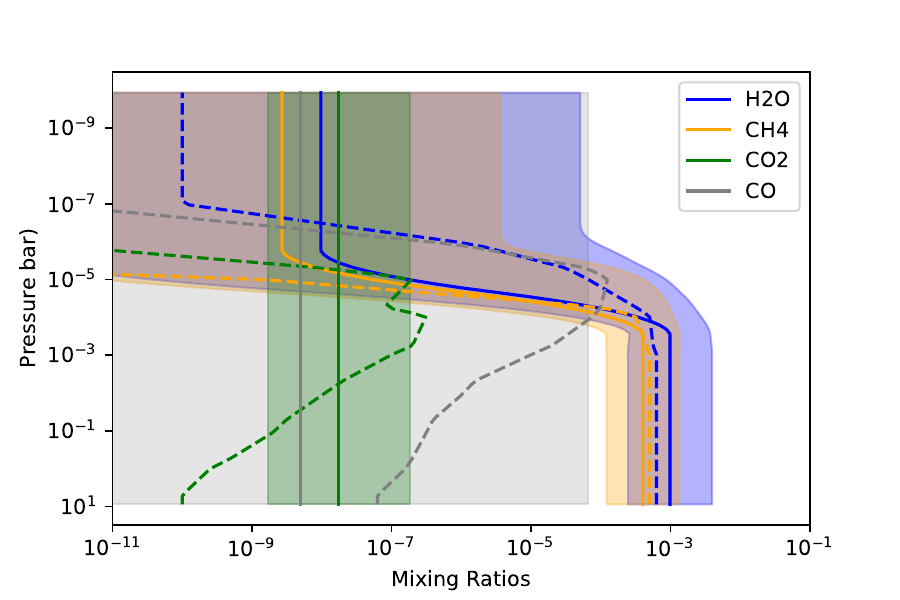}
    \includegraphics[width=0.45\textwidth]{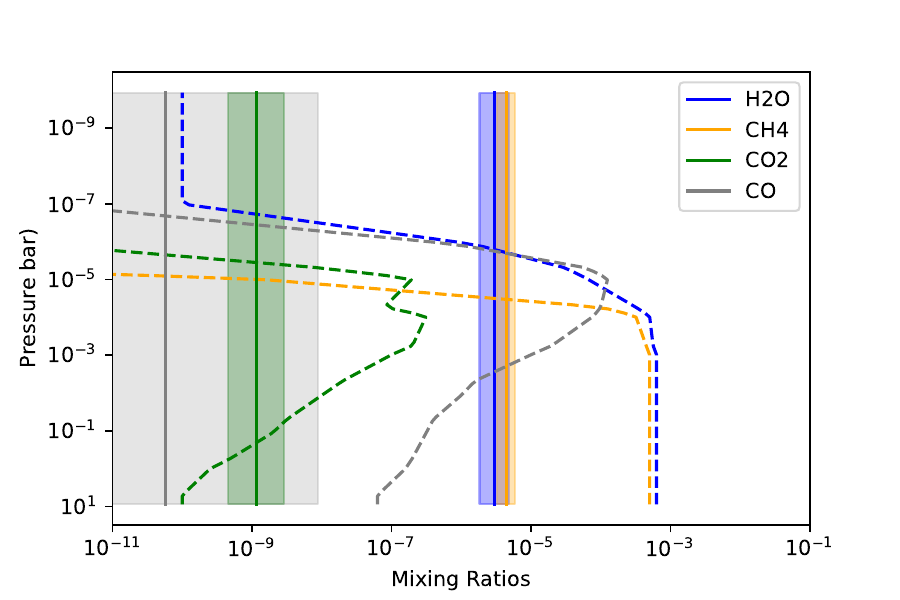}
\caption{
Results of the retrieval for a planet like GJ\,1214\,b. Left: 2-layer profile for H$_2$O and CH$_4$, while CO$_2$ and CO use constant profiles. Right: all molecules are retrieved with constant (1-layer profile) chemistry. Top rows: fitted spectra, middle rows: temperature profiles and bottom rows: chemical profiles of H$_2$O, CH$_4$ and CO$_2$ and CO. For the temperature and the chemical profiles, the dotted lines correspond to the input values. For the 2-layer run we obtain log(E) = 736, while for the 1-layer run we get log(E) = 691 ($\Delta log(E) = 45$).
}\label{fig:GJ1214_retrieval}
\end{figure*}

In the forward model, the particle size $R_{clouds}$ was assumed to be 0.01 microns while $\chi_{clouds}$ was set to $10^{-6}$. We also simplify our problem by assuming the hazes cover the entire atmospheric pressure range. The parameter $Q_0$ describes the type of clouds and is, in this example, fixed to the value of 80 as it can be informed from theoretical models. The bounds of the cloud parameters are chosen to cover a wide range of possibilities: $R_{clouds}$ varies between 0.003 - 1 $\mu m$ and $\chi_{clouds}$ varies between $10^{-15}$ - $10^{-3}$. As with WASP-33\,b, we chose a 3-point thermal profile and apply the same  bounds (490K - 1690K). For these two cases, the full posteriors are presented in Appendix Figure \ref{fig:appendinx_GJ1214b} and Figure \ref{fig:appendinx_GJ1214b_cst}. The fitted spectrum, the chemical profiles and the TP profile are presented in Figure \ref{fig:GJ1214_retrieval}.

We find that for the 2-layer hybrid scenario, all input chemical parameters except the abundance of CO can be recovered. Due to the high opacity, we retrieve weak constraints on the temperature parameters at high pressure, returning large posteriors for $T_{surf}$, $T_1$ and the associated pressures. The cloud parameters are retrieved within the expected values. We notice a strong correlation between the clouds particle size and their abundances.

In the full 1-layer scenario, the solution provides a good fit to the observation, while the retrieved abundances for H$_2$O and CH$_4$ seem to average the real profiles. However, it appears that the temperature profile exhibit large variations, especially for pressures lower than $10^{-2}$ bar, where the true temperature profile is outside the retrieved 1-sigma temperature. The difficulty encountered by the 1-layer scenario in explaining the spectrum is also confirmed by the retrieved posteriors in Appendix Figure \ref{fig:appendinx_GJ1214b_cst} where, in particular, the temperature and pressure of point 2 are offset from the true values and are degenerate with the other parameters. We also find that the radius is not as well retrieved as in the 2-layer hybrid case. In practice, the 1-layer solution would be unlikely to be accepted since the retrieved temperature points tend to push towards values outside reasonable priors.

This result confirms that the isothermal assumption could lead to biases in retrievals of JWST and ARIEL \citep{Rocchetto_biais_JWST}. Additionally, the mixing ratio of around $10^{-10}$ of H$_2$O in the upper atmosphere is too low to be captured by observations given the large haze opacity assumed. For this planet, the detection limit of H$_2$O at this altitude is around $10^{-5}$, correctly interpreted by the large error bars. Also for this example, by using a 2-layer retrieval, correlations in the chemical profiles and detection of cloud layers in transit spectra could inform us about the nature of hazes and clouds. 


\section{DISCUSSION}

\subsection{A physically motivated reason to consider non-constant vertical chemical profiles}

We have shown in the previous sections that simulated atmospheres with a 2-layer chemical profile would induce  spectral features that need to be properly accounted for in retrievals to avoid incorrect conclusions. However, one  could ask whether such family of chemical profiles can be found in exoplanetary atmospheres. We have already demonstrated through the cases of WASP-33\,b and GJ\,1214\,b that chemical profiles with vertical discontinuities could be  important if clouds and hazes are present in the atmosphere.

Additionally,  chemical simulations by \cite{venot_chem_HJ} suggest at least two typical behaviours for chemical profiles in exoplanetary atmospheres of the type HD\,209458b. Some molecules of interest, such as H$_2$O and CO, are predicted to have a constant  mixing ratios as a function of pressure. Others, like NH$_3$ or CH$_4$, are expected to vary with altitude. In the deep atmosphere (generally pressures higher than $1$ bar / $10^5$ Pa) chemical reactions are close to their thermochemical equilibrium values. In the higher part of the atmosphere ($\sim 10^{-4}$ bar / $10$ Pa) photo-chemistry and disequilibrium processes may modify the overall mix by dissociation and creation of atomic species and new molecules. In addition, \cite{Moses_GJ436} investigated the composition of Hot Neptunes like GJ\,436\,b with a wide range of metallicities and the resulting chemical profiles demonstrated complex behaviours. This highlights the need for adapted retrieval techniques. These disequilibrium processes are expected to be more prominent and important in colder atmospheres \citep{Tinetti_ariel}. 

\begin{center}
\begin{figure}[h]
\centering
    \includegraphics[width=0.42\textwidth]{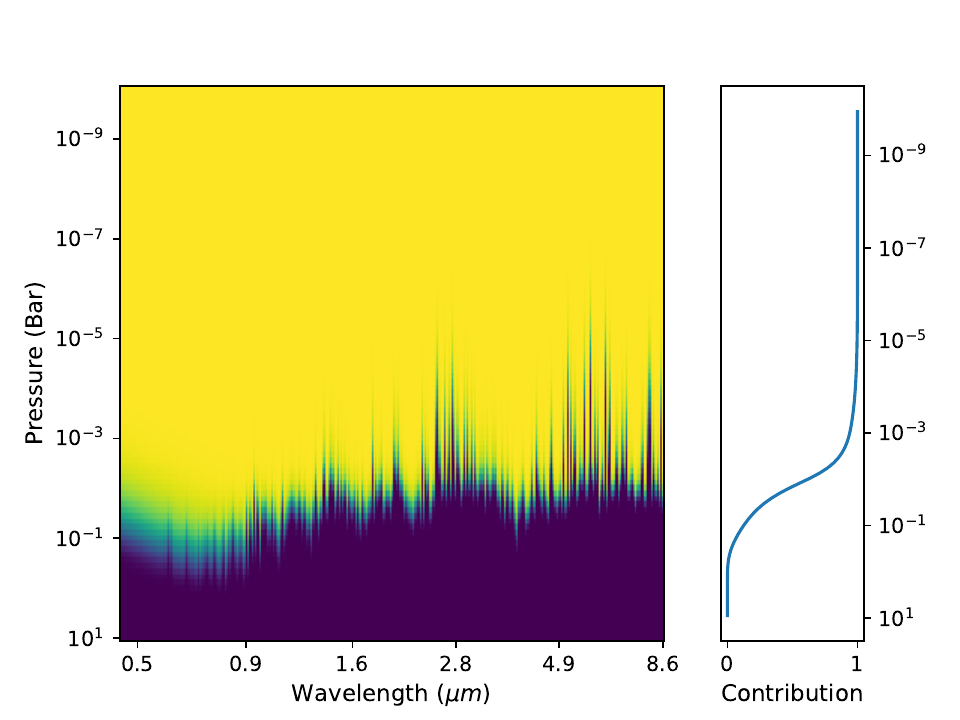}
    \includegraphics[width=0.42\textwidth]{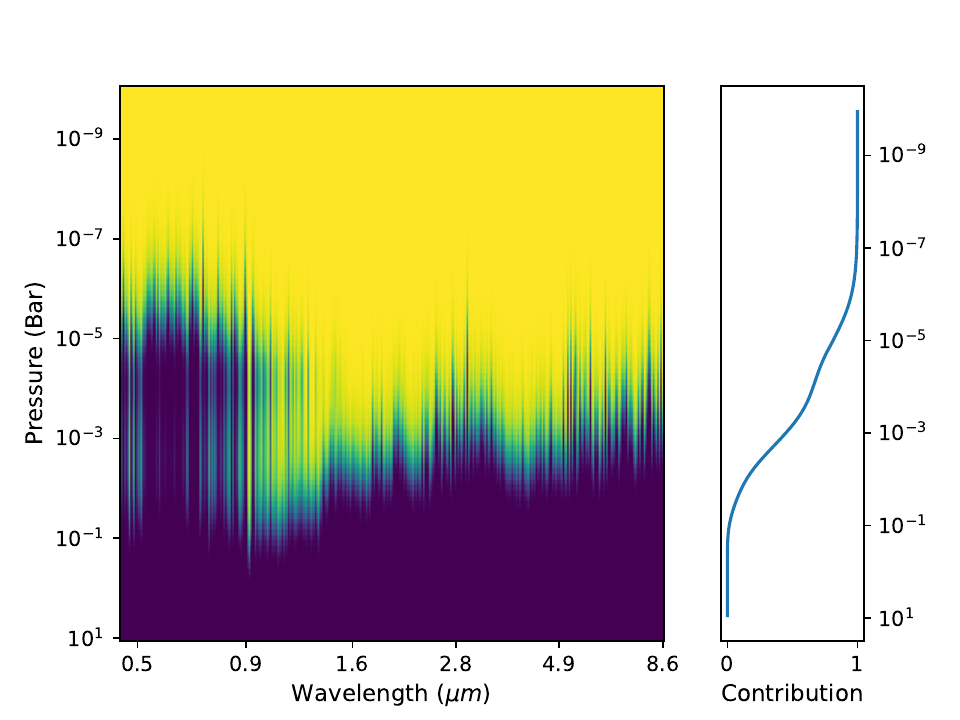}
    \includegraphics[width=0.42\textwidth]{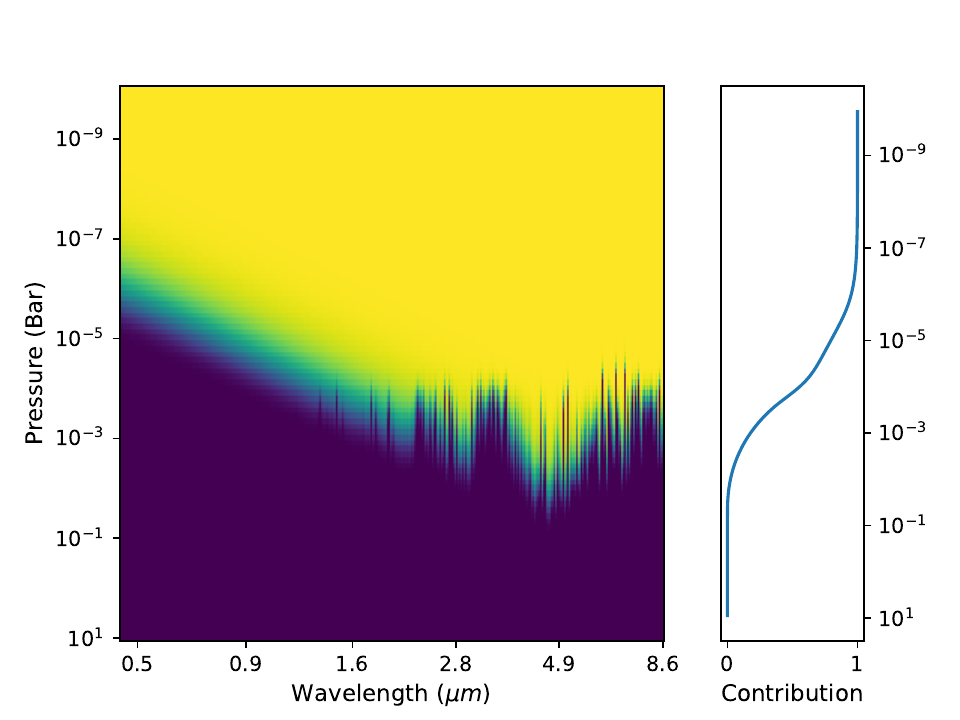}
\caption{
Opacity contribution functions for: Top: a Hot-Jupiter (e.g. HD\,209458\,b); Middle: an Ultra Hot-Jupiter (e.g. WASP-33\,b); Bottom: a Sub-Neptune (e.g. GJ\,1214\,b). In each plot, the left panel  shows the contribution function as function of wavelength (horizontal axis) and the pressure (vertical axis). The right panel is the same function averaged over all wavelengths. For a Hot Jupiter like HD\,209458\,b, the pressures probed range from $1$ bar to $10^{-4}$ bar. For an Ultra-Hot Jupiter of the type WASP-33\,b, the contribution ranges from $10^{-1}$ bar to $10^{-7}$ bar. For a Mini-Neptune planet like GJ\,1214\,b, the contribution function spans the pressures from $10^{-1}$ bar to $10^{-6}$ bar.
}\label{fig:Ariel_contribution}
\end{figure}
\end{center}

Future space instruments should be able to probe roughly between $1$ bar and $10^{-5}$ bar, depending on the composition and temperature of the atmosphere, allowing to constrain chemical models with direct observations. This is showcased in Figure \ref{fig:Ariel_contribution}, where the  plots illustrate the contribution functions and their wavelength dependencies for  planets similar to HD\,209458\,b, WASP-33\,b and GJ\,1214\,b.

\subsection{Should we always use the 2-layer model?}

The increase in complexity in chemical models must be done with care. In some cases, the introduction of additional degrees of freedom comes at the expense of model convergence, i.e: the flexibility of the retrieval should  depend on the quality of the input data. This opens  up the question of model selection. Indeed, should we prefer models with increased flexibility at the risk of increasing model degeneracies and over-fitting, or should we prefer  simpler models but returning only ``acceptable'' fits? 

\begin{center}
\begin{figure}[h]
\centering
    \includegraphics[width=0.49\textwidth]{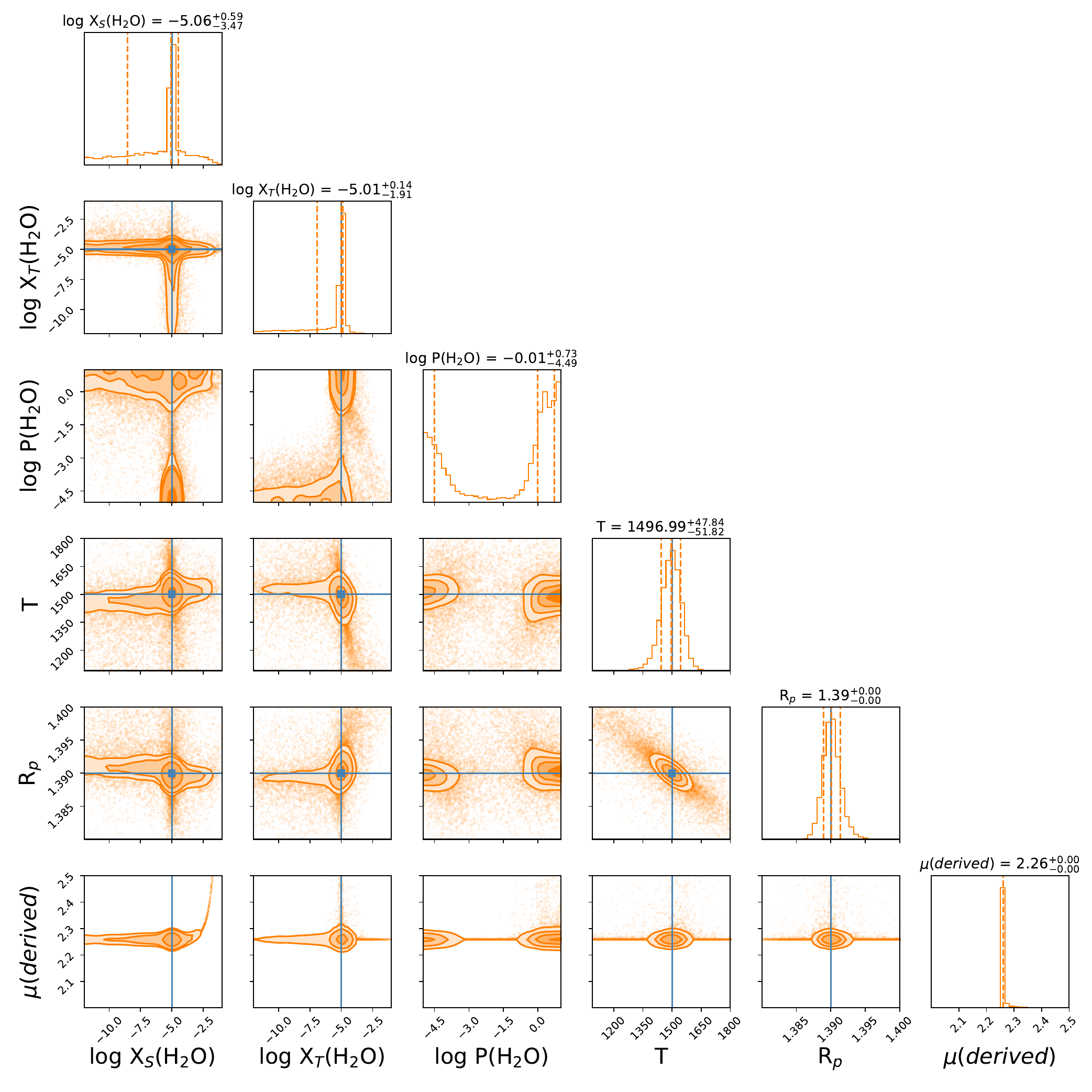}
\caption{Posterior distribution for the retrieval of a constant H$_2$O input profile using the 2-layer model with the Retrieved Pressure Point activated. The model cannot converge as multiple solutions for this point exist. This solution indicates that the number of free parameters is too high and we need to revert to a simpler retrieval.
}\label{fig:H2O_cst_2layer_RPP}
\end{figure}
\end{center}

In the 2-layer case, this issue can be illustrated by the retrieval of a constant input. In section \ref{const with 2 layer}, we disabled the Retrieved Pressure Point to ensure the convergence of the 2-layer retrieval. This choice was justified by the fact that the Input Pressure Point does not exist in constant chemical profiles, making any Retrieved Pressure Point suitable and therefore introducing an intrinsic degeneracy. In Figure \ref{fig:H2O_cst_2layer_RPP} the constant chemical profile used as input is here retrieved with the Retrieved Pressure Point activated (log P(H$_2$O)). The point is however not well constrained and the retrieved abundances become more difficult to interpret. The posteriors are  compatible with a bi-modal solution peaked at pressures where observations are no longer sensitive. 

This example highlights the circumstances under which the model used in the retrieval is too complex.  The issue was solved previously in Figure \ref{fig:cst_2layer_posteriors} by fixing the Retrieved Pressure to an arbitrary value (reduction of the model complexity),  illustrating  that if/when the 2-layer model is too complex for the data, one needs to decrease the number of free parameters and revert back to a simpler chemical parameterisation. This can clearly be seen from the posterior distribution (namely the pressure point divergence).

\section{CONCLUSIONS}

In this paper we have assessed the possibility of constraining the abundance  as a function of altitude of  key chemical species present in exoplanet atmospheres. We have used simulated JWST and ARIEL transit spectra to test whether the data quality of the next generation of space-based instruments will allow for the retrieval of  vertical chemical profiles. 
 The 2-layer model assumed in our paper, while still being  a coarse approximation of the real case, provides an increased level of complexity and flexibility in the interpretation of the data compared to the assumption of constant abundance for each chemical species. To test the validity and usefulness of the model, we included the 2-layer method in the spectral retrieval algorithm TauREx and performed the retrieval of  JWST- and ARIEL-like transit spectra generated by assuming both ad hoc, simplified atmospheric examples and more realistic cases.
 
 We found that the 2-layer retrieval is able to  capture accurately discontinuities in the vertical chemical profiles, which could be caused by disequilibrium processes -- such as vertical mixing or photo-chemistry -- or the presence of clouds/hazes. Our approach should therefore help the removal of the hurdles in interpreting observational constraints that have hindered the confirmation of current chemical models published in the literature. Additionally, the 2-layer retrieval could help to constrain the composition of clouds and hazes by studying the correlation between the chemical changes in the gaseous phase and the pressure at which the condensed/solid phase occurs. This result is particularly important given that clouds/hazes have been detected in half of the currently available exoplanet spectra, but their composition is still elusive as it cannot be inferred directly through remote sensing measurements. 
 
 Future work will extend this analysis to eclipse spectra and explore the need for more complex vertical profiles.  \\
 
 \software{TauREx \citep{Waldmann_taurex2, Waldmann_taurex1}, ArielRad \citep{mugnai_Arielrad}, Multinest \citep{Feroz_multinest}, Corner.py \citep{Foreman-Mackey}}
\newpage

\vspace{5mm}

\noindent\textbf{Acknowledgements}

This project has received funding from the European Research Council (ERC) under the European Union's Horizon 2020 research and innovation programme (grant agreement No 758892, ExoAI), under the European Union's Seventh Framework Programme (FP7/2007-2013)/ ERC grant agreement numbers 617119 (ExoLights) and the European Union's Horizon 2020 COMPET programme (grant agreement No 776403, ExoplANETS A). Furthermore, we acknowledge funding by the Science and Technology Funding Council (STFC) grants: ST/K502406/1, ST/P000282/1, ST/P002153/1, ST/T001836/1 and ST/S002634/1.




\newpage
\bibliographystyle{aasjournal}
\bibliography{main}


\renewcommand{\floatpagefraction}{.9}%
\appendix 

\section*{Planet's parameters used for the forward models}\label{Appendix_parameters}

Parameters used in our forward models for the 3 types of planets: a hot-Jupiter type HD\,209458\,b (\cite{Stassun_planetparam}), an Ultra hot-Jupiter type WASP-33\,b (\cite{Stassun_planetparam}) and a Sub Neptune Type GJ\,1214\,b (\cite{Harpsoe_GJ1214}): 
\[ \begin{array}{|l|c|c|c|}
\mbox{Parameters} & Hot\ Jupiter & Ultra\ Hot\ Jupiter & Sub\ Neptune  \\
\hline
\mbox{$R_s (R_{sun})$} & 1.19 & 1.55 & 0.216 \\
\mbox{$T_s (K)$} & 6091 &  7308 & 3026 \\
\mbox{$R_p (R_{Jupiter})$} & 1.39 & 1.6 & 0.254  \\
\mbox{$M_p (M_{Jupiter})$} & 0.73 & 1.17 & 0.0197 \end{array}\]

\section*{Posteriors of retrievals for a planet with an inverted H$_2$O chemical profile using constant and 2-layer chemical models }\label{Appendix_h2o}

\begin{figure}[H]
\centering
    \includegraphics[width=0.49\textwidth]{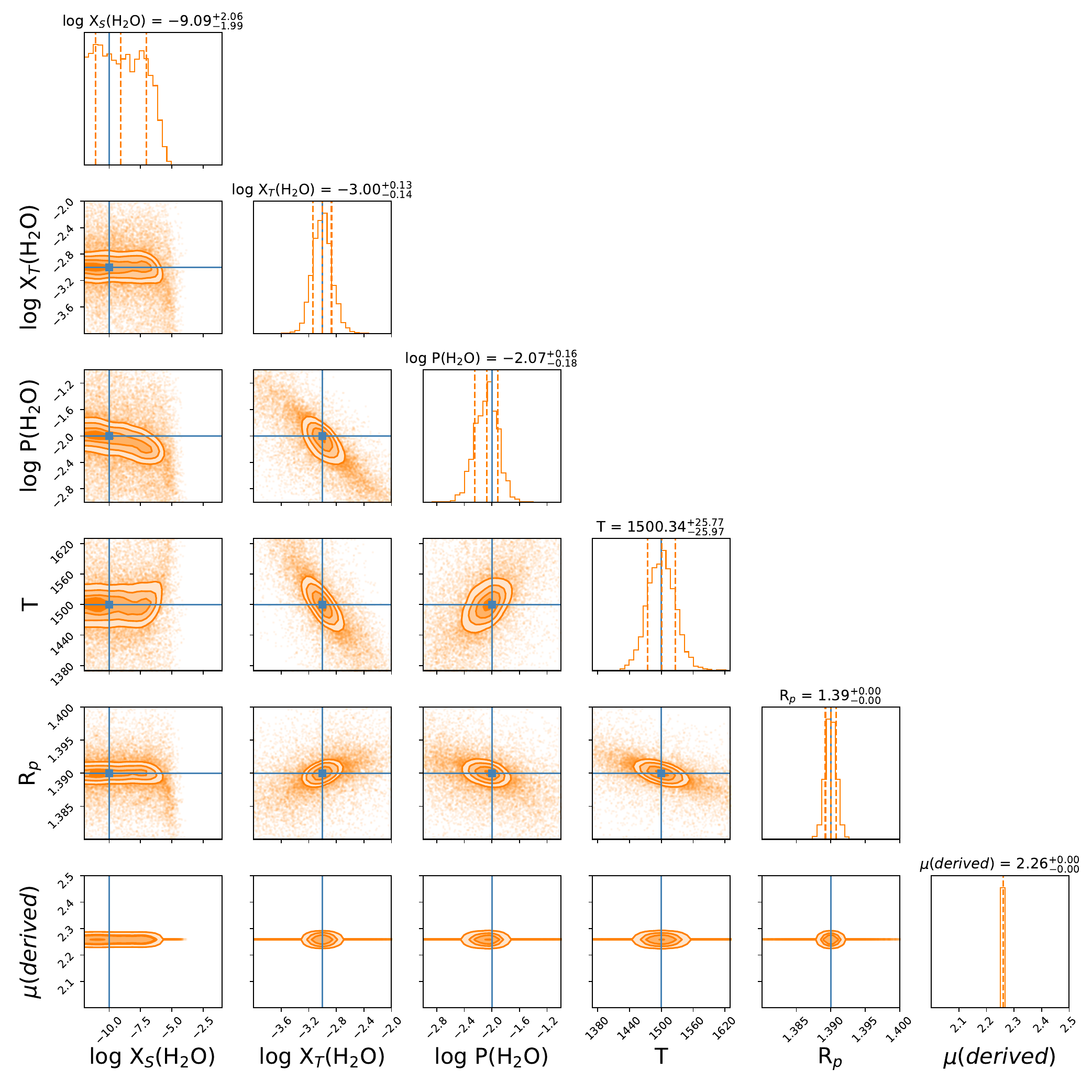}
    \includegraphics[width=0.49\textwidth]{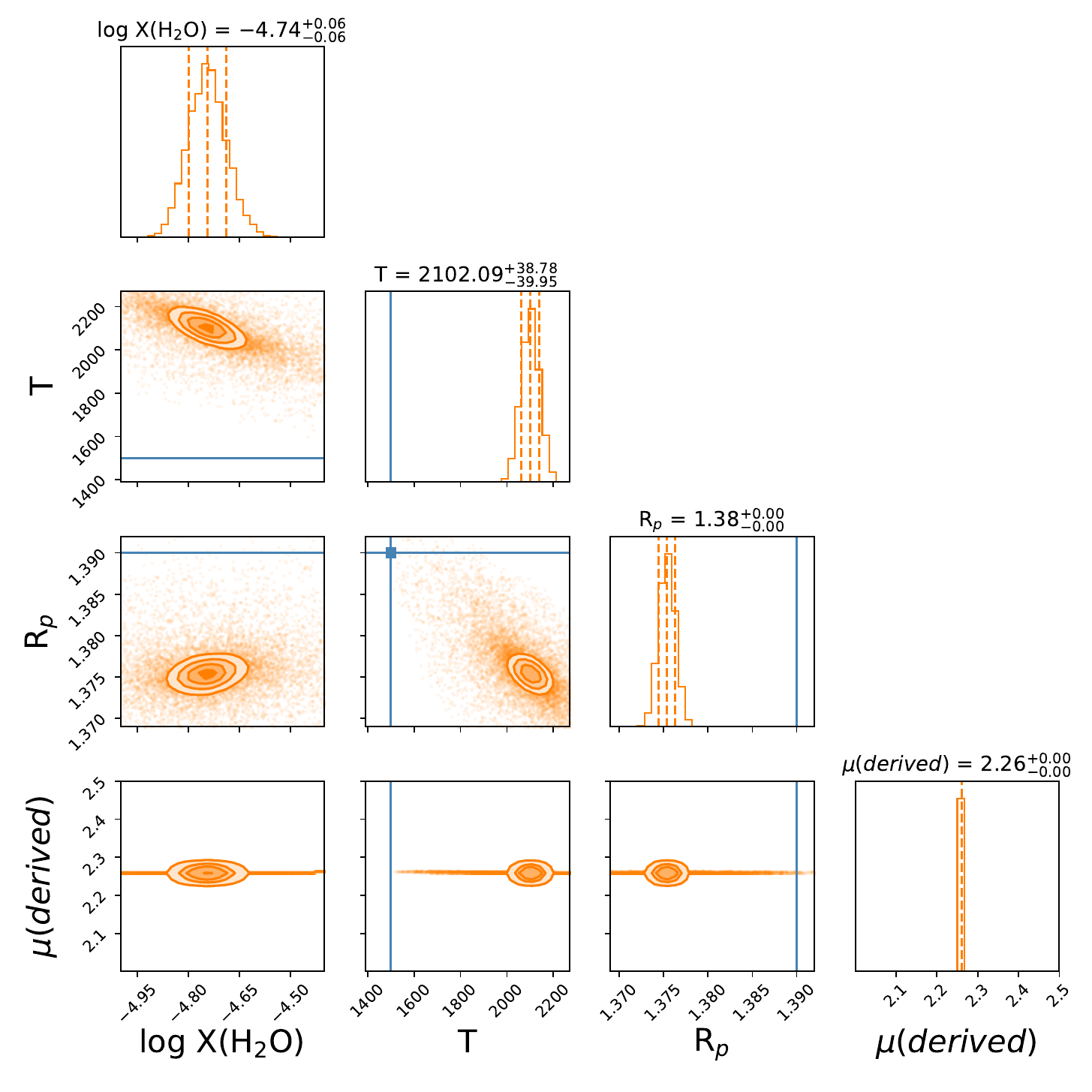}
    \caption{Posteriors of the 2-layer retrieval (left) and the constant retrieval (right) for a simulated ARIEL observation of a planet with an inverted H$_2$O profile. The top layer contains $X_T($H$_2$O$) = 10^{-3}$ for pressures lower than $P_I($H$_2$O$) = 10^{-2}$ bar and the surface layer is depleted with a mixing ratio of only $X_S($H$_2$O$) = 10^{-10}$.}
    \label{fig:appendix_H2O}
\end{figure}

\section*{Posteriors distribution for the retrieval of WASP-33\,b}\label{Appendix_Wasp33b}

\begin{figure}[H]
    \centering
    \includegraphics[width=0.99\textwidth]{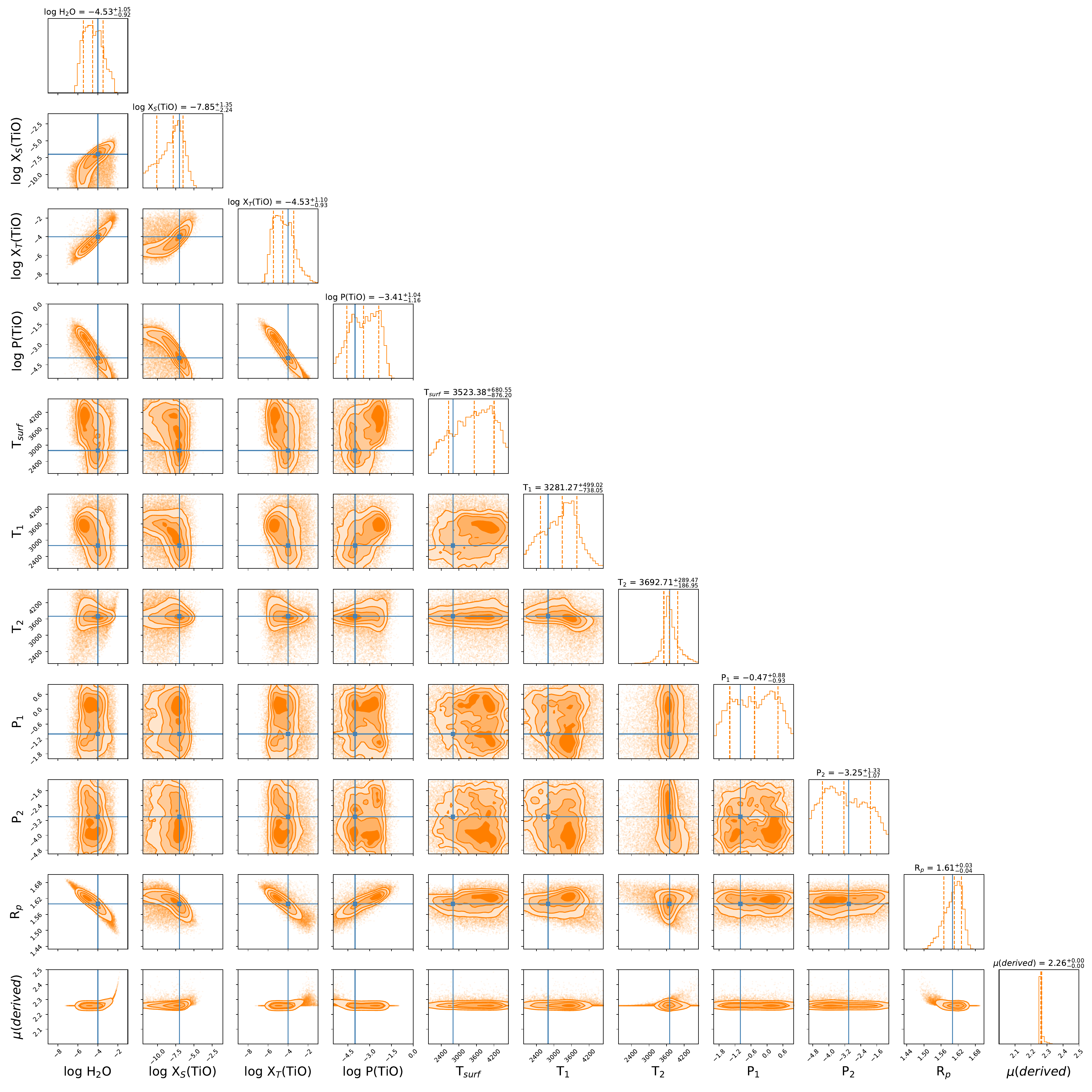}
    \caption{Posteriors distribution for the retrieval of WASP-33\,b. The planet presents constant H$_2$O abundance and a TiO 2-layer profile with a large abundance in the upper atmosphere.}
        \label{fig:appendinx_Wasp33b}
\end{figure}

\section*{Posteriors distribution for the retrieval of GJ\,1214\,b with 2-layer profiles for H$_2$O and CH$_4$ and constant profiles for CO$_2$ and CO}\label{Appendix_GJ1214b}

\begin{figure}[H]
\centering
    \includegraphics[width=0.99\textwidth]{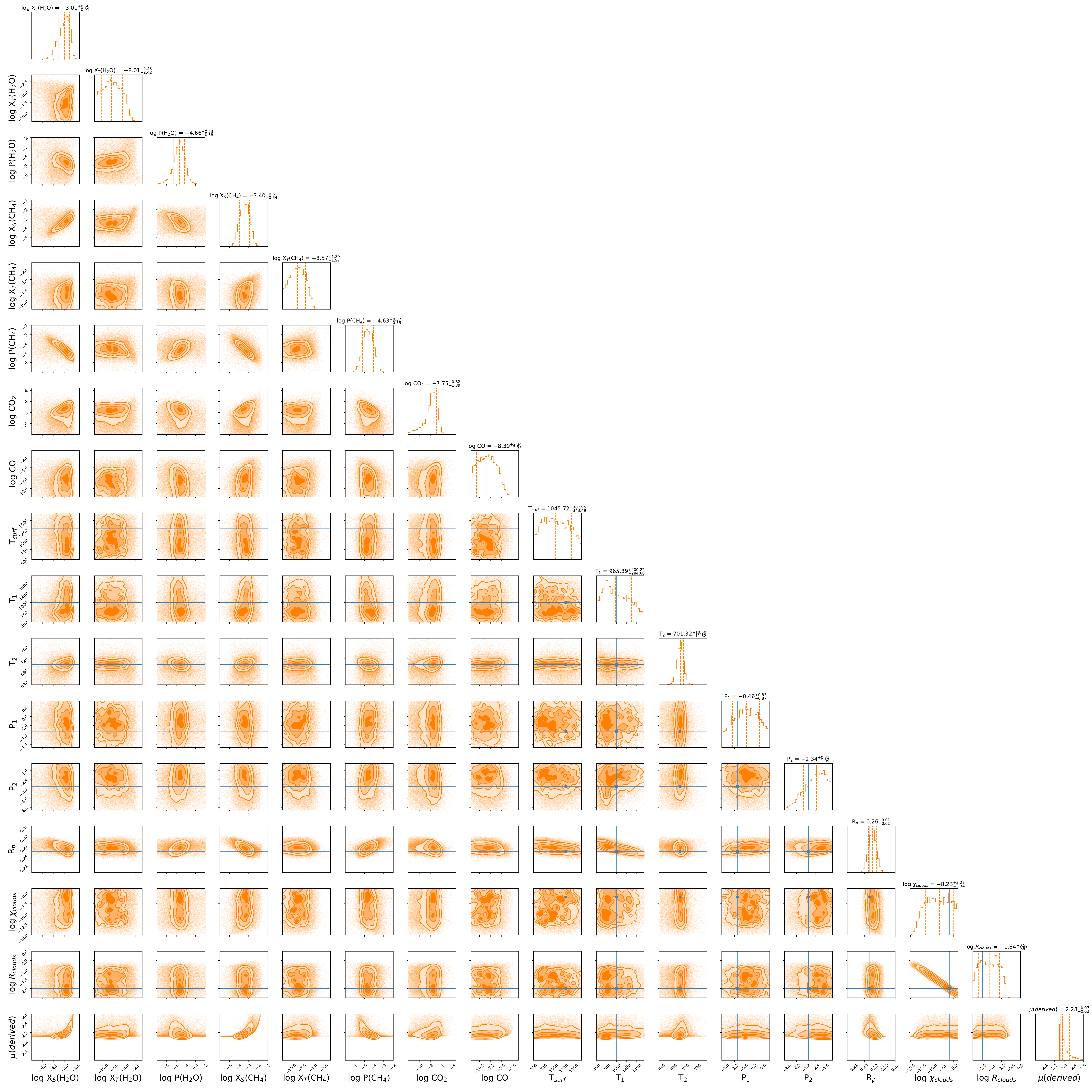}
    \caption{Posteriors distribution for the retrieval of GJ\,1214\,b with hydrocarbon hazes. The planet is retrieved using 2-layer for H$_2$O and CH$_4$. Hydrocarbon hazes are added in the atmosphere to simulate the irradiation of CH$_4$.}
    \label{fig:appendinx_GJ1214b}
\end{figure}

\section*{Posteriors distribution for the retrieval of GJ\,1214\,b in the case of constant chemical profiles}\label{Appendix_GJ1214b_cst}

\begin{figure}[H]
\centering
    \includegraphics[width=0.99\textwidth]{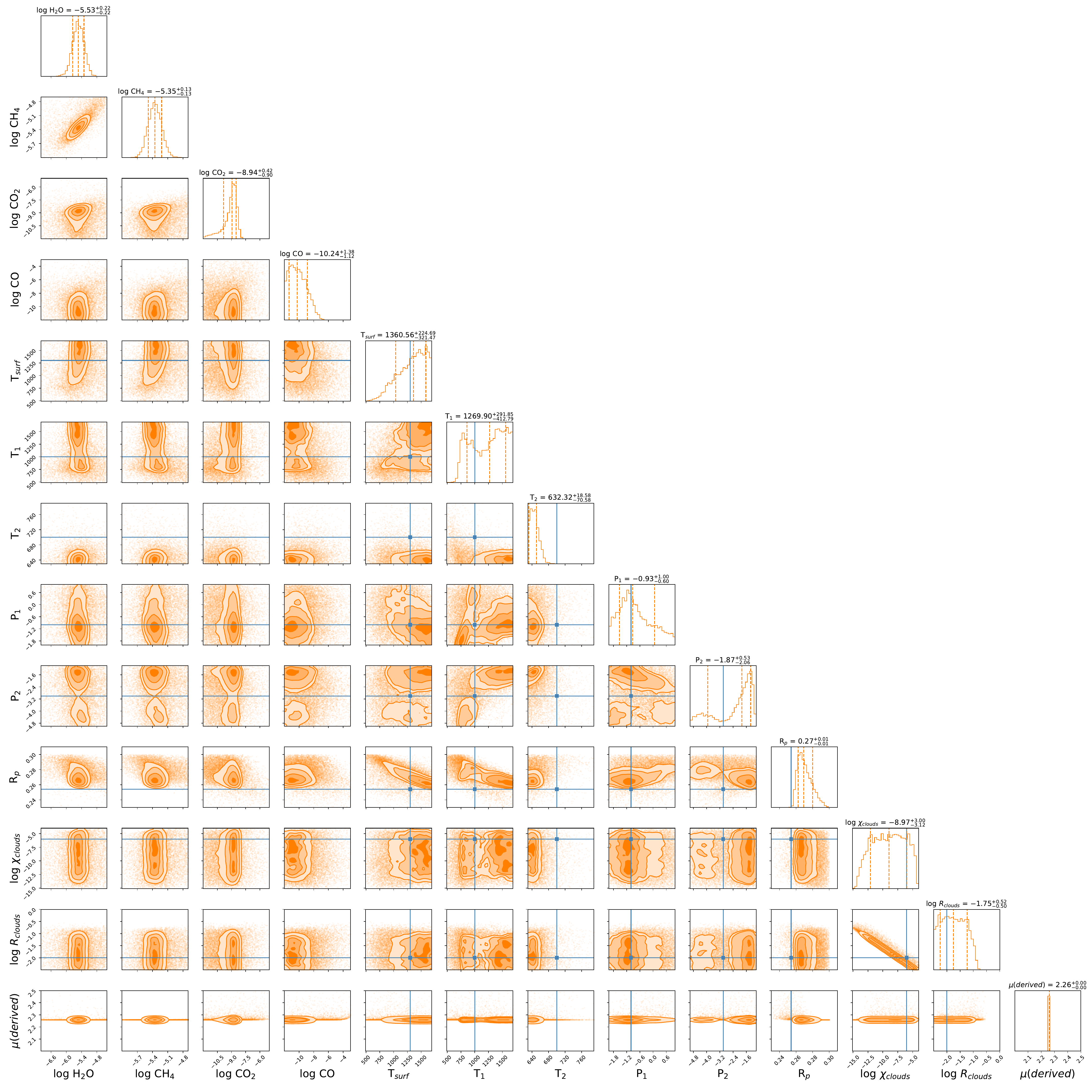}
    \caption{Posteriors distribution for the retrieval of GJ\,1214\,b with hydrocarbon hazes. The planet is retrieved using constant chemistry for all molecules.}
    \label{fig:appendinx_GJ1214b_cst}
\end{figure}


\end{document}